\newcommand{\version}{February 26, 2020}
\documentclass[letterpaper,oneside,11pt,onecolumn,pdftex]{article}
\pdfoutput=1
\usepackage[bindingoffset=0.0cm,textheight=22.6cm,hdivide={*,16.0cm,*}, vdivide={*,22.6cm,*}]{geometry}
\usepackage{amsmath,amsfonts,amssymb}
\usepackage{mathtools}
\usepackage[pdftex]{graphicx}
\usepackage[margin=0.5cm,font=small,labelfont=bf]{caption}
\usepackage[T1]{fontenc}
\usepackage[utf8]{inputenc}
\usepackage{lmodern}

 \usepackage[numbers,square,comma,sort&compress]{natbib}
\usepackage[pdftex,hyperref,svgnames]{xcolor}
\usepackage[pdftex,bookmarksnumbered=true,breaklinks=true,%
colorlinks=true,linktocpage=true,linkcolor=MediumBlue,citecolor=ForestGreen,urlcolor=DarkRed]{hyperref}

\makeatletter
\AtBeginDocument{
  \hypersetup{
    pdfkeywords = {\keywords},
    pdftitle = {\@title},
    pdfauthor = {\@author}
  }
}
\makeatother

\makeatletter

 \@addtoreset{equation}{section}
 \makeatother

\newcommand{\vare}{\varepsilon}

\renewcommand{\th}{\theta}

\newcommand{\rh}{\rho}

  \newcommand{\cL}{\mathcal{L}}

 \newcommand{\cT}{\mathcal{T}}

\newcommand{\ml}{\mathfrak{l}}
\newcommand{\mm}{\mathfrak{m}}
\newcommand{\mn}{\mathfrak{n}}

\newcommand{\nn}{\nonumber}
\newcommand{\eqnref}[1]{Eq. \eqref{#1}}

\newcommand{\kb}{k_{\txt{B}}}
\DeclareMathOperator\erf{erf}

\newcommand{\ct}{c_{\textrm{t}}}

\newcommand{\bt}{\beta_{\textrm{t}}}

\newcommand{\txt}[1]{\textrm{#1}}

\newcommand{\coleq}{\vcentcolon=}
\title{\texorpdfstring{\begin{flushright}
        {\small LA-UR-19-26796}
       \end{flushright}\vspace{2em}}{}%
       Analytic model of the remobilization of pinned glide dislocations: \texorpdfstring{\\}{}%
       including dislocation drag from phonon wind}

\author{Daniel N. Blaschke, Abigail Hunter, Dean L. Preston}

\date{\version}

\newcommand{\keywords}{dislocations in crystals, constitutive behavior, shock waves, drag coefficient}

\graphicspath{{./}{./figures/}}

\begin{document}

\maketitle

\thispagestyle{empty}
\begin{center}
\vspace{-0.3cm}
Los Alamos National Laboratory\\Los Alamos, NM, 87545, USA
\\[0.5cm]
\ttfamily{E-mail: dblaschke@lanl.gov, ahunter@lanl.gov, dean@lanl.gov}
\end{center}

\vspace{1.5em}

\begin{abstract}
In this paper we discuss the effect of a non-constant dislocation drag coefficient on the very high strain rate regime within an analytic model describing mobile-immobile dislocation intersections applicable to fcc polycrystals.
Based on previous work on dislocation drag, we estimate its temperature and pressure dependence and its effects on stress-strain rate relations.
In the high temperature regime, we show that drag can remain the dominating effect even down to intermediate strain rates.
We also discuss the consequences of having a limiting dislocation velocity, a feature which is typically predicted by analytic models of dislocation drag,
but which is somewhat under debate because a number of MD simulations predict supersonic dislocations.
\end{abstract}

\tableofcontents
\newpage

\section{Introduction}
\label{sec:intro}
%%%%%%%%%%%%%%%%%%%%%%%%%%%%%%%%%%%%%%%%%

To accurately predict material response for explosively-driven systems, hypervelocity impacts, and more generally, material deformation and failure under shock wave loading, reliable material strength models, or plastic constitutive models, applicable at strain rates of $\sim10^5 \mathrm{s}^{-1}$ and higher are essential.
In this high rate loading regime, intersections of non-coplanar, attractive mobile and immobile (forest) dislocations are no longer the dominant rate-controlling mechanism, and hence do not dominate the work hardening behavior seen in the overall material response.
Rather, drag effects on dislocation motion due to phonon wind are dominant.
At high strain rates, dislocations moving through the crystal experience a ``friction'' due to their interaction with phonons.
At dislocation velocities of a few percent sound speed this friction or drag coefficient $B$ is known to be roughly constant~\cite{Nadgornyi:1988}, but at higher speeds its functional form has historically been poorly understood.
This is in part due to the difficulty in accessing these regimes experimentally, especially with non-destructive and \textit{in situ} methods.
Experiments to directly measure dislocation drag from phonon wind have been done, however only at low dislocation velocities (i.e. $\le1\%$ transverse sound speed), ambient pressure, and a small number of temperatures up to 300K \cite{Nadgornyi:1988}.
Similar experiments for varying pressures and high temperatures have yet to be done, and measurements at high dislocation velocities are not possible with current techniques.
Without a clear understanding of how the drag coefficient varies with temperature, pressure, and in particular velocity, important questions about how dislocations move under extreme loading conditions remain.
For example, whether dislocations in metals can reach supersonic speeds or are limited by the lowest shear wave speed (or any other sound velocity) is still under debate.
Notably, numerous MD simulations suggest it is possible~\cite{Rosakis:2001,Li:2002,Olmsted:2005,Jin:2008,Gilbert:2011,Ruestes:2015}, but compare dislocation velocities to the sound speeds of the undeformed crystal.
% and results are somewhat sensitive to the choice of interatomic potential used in a simulation.
Furthermore, there has been no experimental validation of such behavior in metals\footnote{
There has been experimental evidence for supersonic dislocations in a plasma crystal~\cite{Nosenko:2007}, however.}.
There has also been recent discussion on interpreting those results in the context of line tension and dislocation shape~\cite{Blaschke:2017lten},
and other recent discussions on supersonic dislocations can be found in \cite{Markenscoff:2009,Pellegrini:2010,Pellegrini:2017} and references therein.
What is clear from MD results, however, is that dislocation drag is greatly enhanced in the high velocity regime.

In this work, a model for the dependence of the dislocation drag coefficient developed using first principle calculations (i.e. derived from fundamental theoretical calculations) is incorporated into a plastic constitutive model, which has generalized the standardly used low strain rate relationship up to strain rates of roughly  $10^{10} \mathrm{s}^{-1}$ for fcc polycrystals, allowing for the effects of dislocation drag on the overall material response in the high-rate loading regime to be studied.

It is well known that metals and alloys exhibit a significant increase in strain rate sensitivity at rates $\sim 10^{4-5} \mathrm{s^{-1}}$ \cite{Meyers:1994}.
This behavior is usually interpreted as a change in the rate-controlling mechanism from thermally-activated glide at low rates, typically described by a van 't Hoff-Arrhenius equation \cite{Seeger1:1954, Seeger2:1954, Seeger3:1954, Seeger:1955, Follansbee:1988, Hoge:1977, Steinberg:1989, Zerilli:1987, PTW:2003}, to dislocation drag at high rates.
In the thermally activated regime, work hardening in metals and alloys has been shown to be predominated by pairwise interactions of dislocations \cite{Friedel:1964,Madec:2002}, an approximation we adopt here also.
As two attractive dislocations near each other, a dislocation junction can form, pinning glissile dislocations.
The pinned (glissile) dislocation segments cannot continue to move, and thus accommodate the loading conditions via plasticity, until the dislocation intersection 
is broken.
At relatively low applied stresses or strain rates, the applied loading is not enough to break these dislocation intersections at $T=0$.
However, at finite temperatures ($T>0$), atomic oscillations yield local stress fluctuations that can assist the applied loading in moving mobile dislocations past forest dislocations and other short-ranged obstacles \cite{Friedel:1964,Becker:1925,Orowan:1934a,Orowan:1934b,Orowan:1934c}, thus resulting in node dissociation.
The inverse of the node lifetime is generally calculated according to the van 't Hoff-Arrhenius rate equation \cite{Arrhenius:1889,vantHoff:1896},
\begin{align}
 \Gamma^{-1}=f_a\txt{exp}\left(-\frac{E_a}{\kb T}\right)
 \,,\label{eq:Arrhenius}
\end{align}
where $E_a$ is the activation (Gibbs free) energy, $f_a$ is an ``attempt frequency'', $\kb$ is the Boltzmann constant, and $T$ is the temperature.
While van 't Hoff-Arrhenius thermal activation theory yields a good description below strain rates of $10^{4-5}$s$^{-1}$ where the dislocation drag coefficient can be neglected, it breaks down at higher strain rates.
This is because the van 't Hoff-Arrhenius equation \eqref{eq:Arrhenius} is valid only when the height of the potential barrier is large compared to the thermal energy, ${E_a}/{\kb T}\gg1$.
Since $E_a$ decreases with increasing applied stress (because the resulting strain rate must increase), \eqnref{eq:Arrhenius} breaks down at sufficiently high stresses.
Many rate-dependent plastic flow models \cite{Seeger1:1954, Seeger2:1954, Seeger3:1954, Seeger:1955, Follansbee:1988, Hoge:1977, Steinberg:1989, Zerilli:1987} that rely on the van 't Hoff-Arrhenius model \cite{Arrhenius:1889}, exhibit a progressive drop in fidelity and in some cases explicit failure at high stresses.
Many current models are combinations of sub-models for the thermally-activated regime (often based on van 't Hoff-Arrhenius, but not always) and the drag-dominated regime plus a prescription for the transition between the sub-models \cite{PTW:2003,Langer:2010,Austin:2011,Yanilkin:2014,Barton:2011}.

Several different approaches for modeling the drag dominated regime have been developed, including different assumptions about the drag coefficient, $B(v, P, T)$, itself.
Some more phenomenological models do not explicitly include the dislocation drag coefficient at all, such as the one in Ref.~\cite{PTW:2003} which relied on experimental shock data for their model parameters in order to make predictions in the high stress regime.
More recent approaches \cite{Krasnikov:2010,Barton:2011,Luscher:2016,Austin:2018,Olmsted:2005,Marian:2006,Cho:2017,Hunter:2015} aim at providing a more accurate description of the high stress regime based on the microscopic physics of dislocation mobility.
However, one of the main obstacles encountered by these ``microscopic'' strength models has been the uncertainty as to how $B$ behaves at high velocities, temperatures, and pressures.
Most commonly, some or all of the dependencies of the drag coefficient are neglected, ranging from conservative choices like $B=\textrm{const.}$
(see e.g. \cite{Kuksin:2008,Hansen:2013,Hunter:2015,Borodin:2015}) to $B\sim\sqrt{v}$ above some threshold velocity \cite{Olmsted:2005,Marian:2006,Cho:2017}, to ``relativistic factors'' $B\sim1/(1-v^2/v_\txt{crit}^2)^m$ with a limiting (critical) velocity $v_\txt{crit}$ and a range of powers $1/2\le m\le 4$ \cite{Krasnikov:2010,Barton:2011,Luscher:2016,Austin:2018}.
In these references, the pressure and density dependence of $B$ is largely ignored (except for the ``relativistic factors'' whose limiting velocity depends on the shear modulus).
Whenever temperature dependence is considered, it is assumed to be linear, sometimes including a small quadratic correction \cite{Barton:2011}.
The latter assumption has its roots in the linear temperature dependence of the Debye phonon spectrum in its high temperature expansion~\cite{Alshits:1992,Blaschke:BpaperRpt}.
However, none of the velocity dependencies mentioned above are well motivated from first principles, but are rather assumptions proposed ad hoc.
For example, $B\sim\sqrt{v}$ is based on Eshelby's  arguments~\cite{Eshelby:1956} for screw dislocations in an isotropic continuum supplemented by an anisotropic dispersion relation and cannot be used for any other dislocation character.
This may have been the best estimate when it was put forward in the 1950s, but today we know it gives incorrect high velocity behavior even for screw dislocations \cite{Blaschke:2018anis}.
A relativistic factor with power $m=1$ was introduced in Refs. \cite{Gillis:1969,Clifton:1971} in order to introduce a limiting velocity using a simple analytic form to account for the divergence in self energy of dislocations in the isotropic limit \cite{Weertman:1961}.
Numerous recent authors have copied this form \cite{Austin:2011,Barton:2011,Luscher:2016,Austin:2018} and some have changed the power $m$ to better match their simulations; see e.g. \cite{Krasnikov:2010} and references therein.

Recently \cite{Hunter:2015} developed an analytic model of mobile-immobile dislocation intersection that generalizes the standard low plastic strain-rate relation $\dot{\varepsilon}_\txt{p}=\dot{\varepsilon}_0 \exp (-E(\sigma)/\kb T) $ up to strain rates of roughly $10^{10} \mathrm{s}^{-1}$ (applicable to fcc polycrystals), which they termed the kinetic equation.
The kinetic equation, based on mean-first-passage-time (MFPT) theory, provides a plastic strain rate - flow stress relationship that is applicable without ``connecting'' models for different strain-rate regimes.
Although the model framework is valid for a velocity-, density-, and temperature-dependent drag coefficient, $B(v,\rho,T)$, the focus in \cite{Hunter:2015} was dislocation intersection node dissociation, hence for simplicity $B$ was taken to be a constant.
In addition to the development of the kinetic equation, Blaschke \textit{et al.} developed a new functional form for $B$ based on first principles calculations of elastic scattering of phonons by a moving dislocation via linear response theory \cite{Blaschke:BpaperRpt,Blaschke:2018anis,Blaschke:2019fits,Blaschke:2019Bpap}, thereby generalizing earlier work \cite{Alshits:1992}.
This model for the drag coefficient focuses primarily on the dependence of $B$ on the dislocation velocity, and shows that the 
asymptotic behavior of $B$ in the isotropic limit indeed exhibits a ``relativistic'' form with $m=1/2$ and $v_\txt{crit}=\ct$ (the transverse sound speed) \cite{Blaschke:BpaperRpt,Blaschke:2019Bpap}.
Such a limiting dislocation velocity (not surprisingly) leads to a limiting strain rate $\dot\vare_\txt{max}$ at a given dislocation density.
The impact this has on the overall material response has not yet been studied.
A generalization of this model to fully include temperature and density dependence of $B$ is currently in progress~\cite{Blaschke:2019};
However the only missing piece in this regard is the temperature and density dependence of the ratio of polynomials of second (SOEC) and third order (TOEC) elastic constants, as discussed below in Section~\ref{sec:dragcoeff}.
Therefore, in approximating this ratio to be constant with temperature and density, it is already possible to derive a first estimate of $B(v,\rho,T)$.

In this work, we have integrated the models of Hunter and Preston \cite{Hunter:2015} and Blaschke \textit{et al.} \cite{Blaschke:BpaperRpt,Blaschke:2018anis, Blaschke:2019fits,Blaschke:2019Bpap} to investigate the impact the drag coefficient, $B(v, \rho, T)$, has on the overall material response, including the evolution of the mobile and immobile dislocation density populations.
For completeness, Section \ref{sec:summary} starts with a brief review of this plastic constitutive model as presented in Ref.~\cite{Hunter:2015}.
The kinetic equation as formulated in Ref.~\cite{Hunter:2015} presents the plastic strain rate as a function of the flow stress.
However, for numerical simulations involving plastic deformation, the flow stress as a function of strain rate is generally required.
Hence, in Ref.~\cite{Hunter:2015} the inverse kinetic equation is also presented (and reviewed in Section \ref{sec:summary}).
In order to include the drag coefficient of Blaschke \textit{et al.} \cite{Blaschke:BpaperRpt,Blaschke:2018anis, Blaschke:2019fits,Blaschke:2019Bpap}, an analytical approximation for dislocation drag as a function of local stress, material density, and temperature,  $B(\sigma,\rho,T)$, valid in the isotropic limit for ambient and high temperatures must be developed.
This is presented in Section \ref{sec:dragcoeff}, and is based on previous work \cite{Blaschke:BpaperRpt,Blaschke:2019Bpap}.
We also note that a generalization of the original model developed by Blaschke \textit{et al.} \cite{Blaschke:2017lten,Blaschke:2018anis,Blaschke:2019fits} included a dependence on the dislocation line character (edge vs. screw), which has been shown to have interesting effects on dislocation drag --- and hence the stress versus strain rate relations.
Since the Hunter-Preston model evolves total dislocation densities, the dislocation character itself is not resolved.
The approximation presented in Section \ref{sec:dragcoeff} averages over dislocation character types so that the dislocation drag model also does not resolve dislocation character dependence, and thus can be integrated with the Hunter-Preston model, as is done in Section \ref{sec:newmodel}.
Generalizations of the model to include the dislocation character dependence are left to future work.
Finally, in Section \ref{sec:results}, we utilize the model to investigate the impact
of the velocity dependence of $B$ on the plastic flow, particularly at high strain rates.

%%%%%%%%%%%%%%%%%%%%%%%%%%%%%%%%%%
\section{The kinetic equation with constant dislocation drag}
\label{sec:summary}
%%%%%%%%%%%%%%%%%%%%%%%%%%%%%%%%%%%%%%%%%

In this section we briefly summarize the main results of the Hunter-Preston model found in Ref. \cite{Hunter:2015}, which includes the formulation of the kinetic and inverse kinetic equations that utilize a constant dislocation drag coefficient in the drag dominated regime.
As mentioned previously, a MFPT framework was developed for intersecting dislocations applicable to high stresses and strain rates in Ref.~\cite{Hunter:2015} in order to generalize the typical thermal activation van 't Hoff-Arrhenius model.
The main result of this approach is the more general form for the remobilization time $t_r$ (replacing $\Gamma$ in \eqnref{eq:Arrhenius}), which in the low stress limit reduces to the van 't Hoff-Arrhenius form.
Consider a crystal with total dislocation density $\rho_\txt{tot}=\rho_m+\rho_i$, decomposed into mobile and immobile (or network/forest) dislocations.
Within this crystal there exist many pinned mobile dislocation segments.
Taking into account the probability that such a segment is formed, the time for it to remobilize (sometimes also 
called the wait time) is as follows:
\begin{align}
 t_r &= \frac12\left[1+\erf(A)\right]t_B\int\limits_0^\infty\exp\left[-\frac{t_B}{\tau}\int\limits_0^z\frac12\left\{1-\erf\left[A\left(1-\frac{\hat\sigma}{\sigma_c}\left(1-e^{-z'}\right)\right)\right]\right\}dz'\right]dz
 \,,\nn\\
 A&=\sqrt{\frac{E_0}{\kb T}}
 \,,\label{eq:deftr}
\end{align}
where $\hat\sigma\coleq(\sigma-\sigma_b)\th(\sigma-\sigma_b)$ denotes the effective applied stress, i.e. with the so-called back stress $\sigma_b$ subtracted from the applied stress $\sigma$ if $\sigma\ge\sigma_b$ and zero otherwise (hence the step function $\th(\sigma-\sigma_b)$).
Additional parameters in this equation are the activation energy at zero stress $E_0$, and $\tau\approx10^{-13}$s which represents a typical time scale for force fluctuations which is smaller than but on the order of the inverse maximum acoustic phonon frequency~\cite{Hunter:2015}.
In addition, there is a stress independent time scale for bow-out of the dislocation segment around the pinning point, $t_B$, henceforth called the bow-out time.
Finally, there is the critical shear stress required to dissociate the node, $\sigma_c$.
These quantities were formulated in Ref. \cite{Hunter:2015} using the approximation for the mean line tension in the isotropic limit\footnote{
This expression for the average dislocation line tension follows from assuming an isotropic solid with Poisson ratio of $\nu=1/3$, see e.g. Ref. \cite[p.~176]{Hirth:1982}, and subsequently averaging over the (character) angle $\vartheta$ between dislocation line direction and Burgers vector.
The more recent model of Ref. \cite{Szajewski:2019phil} leads to the same expression.
},
\begin{align}
 \cT &\approx \frac{5}{16\pi}Gb^2\ln\left(\frac{\cL_i}{b}\right)
 \,,\label{eq:linetension}
\end{align}
where $G$ denotes the shear modulus, $b$ is the Burgers vector magnitude, the mean distance between immobile dislocations $\cL_i$ is related to their density $\rho_i$ as $\cL_i\approx1/\sqrt{\rho_i}$, and the dislocation core cutoff was assumed to be of the same order as the Burgers vector, i.e. $r_0\approx b$.
It follows that $\sigma_c$, $\sigma_b$, $t_B$, and $E_0$ can be written as:
\begin{align}
 \sigma_c &= \frac{2\cT\phi_c}{b\cL_i}
 \,,&
 \sigma_b &= \frac{g_b b\, G}{\cL_i}
 \,,\nn\\
 t_B &= \frac{B\cL_i^2}{\pi^2\cT}
 \,, &
 E_0 &= \kappa \phi_c^2\cT b
 \,, \label{eq:estimates}
\end{align}
where $B$ denotes the dislocation drag coefficient which was assumed to be constant and of the order of $0.05$mPa\,s for copper in Ref. \cite{Hunter:2015}, and whose generalization to $B(\sigma,\rho,T)$ will be discussed in Section \ref{sec:dragcoeff} below.
The critical bow-out angle, which is the angle between the initial straight line direction and the tangent to the dislocation line at the pinning point at the time of node dissociation (i.e., at $\sigma_c$), was derived in Ref. \cite{Hunter:2015} to be $\phi_c=\frac{8\pi}{5}k$ with model parameter $k\approx1/10$ extracted from large-scale 3D dislocation dynamics (DD) simulations for fcc metals carried out in Ref.~\cite{Madec:2002}.
Thus, $\phi_c\approx1/2$, and additional model parameters in \eqref{eq:estimates} are $g_b=1/5$ and $\kappa=1$.

We note that in weak external stress fields, thermal fluctuations can induce the reverse motion of mobile dislocations.
In this case, the mean remobilization time for negative (reverse) motion is obtained by simply reversing the sign of the effective applied stress.
Since the mean dislocation velocity and the strain rate are inversely proportional to $t_r$, reverse dislocation glide due to thermally-driven force fluctuations must be taken into account at very low stresses in order to avoid (unphysical) non-zero strain rates at zero stress.
We write the effective mean remobilization time as:
\begin{align}
 \frac1{t_r^\txt{eff}} &= \frac1{t_r(\hat\sigma)} - \frac1{t_r(-\hat\sigma)}
 \,.
\end{align}

\paragraph{Low stress limit:} The mean remobilization time integrations can be carried out in the low stress limit, i.e., $\hat\sigma/\sigma_c\ll1$.
In this case, the factor $e^{-z'}$ can be dropped in the error function within \eqnref{eq:deftr}.
In addition, typical values for $A=\sqrt{{E_0}/{(\kb T)}}$ are a few times unity, and depending on temperature and density of forest dislocations ranging from roughly 2 to 20.
Hence, $A(1-\frac{\hat{\sigma}}{\sigma_c})$ is sufficiently large that the error functions themselves can be approximated by their asymptotic expansions, $\erf(x)\approx1-\frac{1}{\sqrt{\pi}x}e^{-x^2}+\ldots$ as $x\to\infty$.
With these approximations, the integrals over $z'$, $z$ in \eqref{eq:deftr} can be carried out explicitly leading to
\begin{align}
 \frac{t_r}{\tau} &\approx \left[1+\erf(A)\right]\sqrt{\pi}A\left(1-\frac{\hat\sigma}{\sigma_c}\right)\exp\left[A^2\left(1-\frac{\hat\sigma}{\sigma_c}\right)^2\right]
 \,, \label{eq:trlowstress}
\end{align}
whose zero stress limit becomes $\lim_{\hat\sigma\to0}t_r/\tau = \left[1+\erf(A)\right]\sqrt{\pi}A e^{A^2} < \infty$.
Since $\frac{\hat{\sigma}}{\sigma_c}$ is much smaller than 1, this term can be dropped from the factor $1-\frac{\hat{\sigma}}{\sigma_c}$ and $(\frac{\hat{\sigma}}{\sigma_c})^2$ can be dropped from the exponential term.
Based on the aforementioned values of A, $(1+\erf{A})/2\approx 1$.
Substituting for $A$, the remobilization time becomes:
\begin{align}
t_r^{-1} = \frac{1}{2\tau}\left(\frac{\kb T}{\pi E_0}\right)^{\frac{1}{2}}\exp{\left\{-\frac{E_0}{\kb T}\left(1-2\frac{\hat{\sigma}}{\sigma_c}\right)\right\}}
 \,. \label{eq:LStresslimit}
\end{align}
The low stress limit does indeed recover the van 't Hoff-Arrhenius exponential, but with an additional $\sqrt{\kb T/E_0}$ prefactor.
Preliminary results presented in Ref. \cite{Hunter:2015} for copper showed that this prefactor did not produce a pronounced difference in the low stress behavior in comparison to the standard van 't Hoff-Arrhenius model.
As mentioned above, reverse dislocation glide must be considered at very low stresses.
Employing the effective mean remobilization time, the low stress limit can be written as:
\begin{align}
 \lim_{\hat\sigma\to0}t_r^\txt{eff}/\tau
%  = \left[1+\erf(A)\right]\frac{\sqrt{\pi}}{4A} e^{A^2}\frac{\sigma_c}{\hat\sigma} \to \infty 
 \approx\frac{\sqrt{\pi}}{2A} e^{A^2}\frac{\sigma_c}{\hat\sigma} \to \infty
 \,. \label{eq:trlimit}
\end{align}
Note that as $\hat\sigma$ increases, $t_r^\txt{eff}$ tends to $t_r$, but as $\hat\sigma\to\sigma_c$ the approximations leading to \eqnref{eq:trlowstress} break down, i.e. the divergence in $1/t_r(\hat\sigma=\sigma_c)$ is merely an artifact.

\paragraph{High stress limit:} Similarly, the mean remobilization time can be determined for very high stresses, i.e., $\hat\sigma/\sigma_c\gg1$.
With increasing stress, $(1-e^{-z'})\approx z'$ becomes a good approximation, leading to an integral that can be solved analytically within \eqnref{eq:deftr}.
Additional approximations for the remaining integral over $z$ outlined in Ref. \cite{Hunter:2015} lead to the following high stress limit:
\begin{align}
 \frac{t_r(+\hat\sigma)}{\tau} &\approx \frac{t_B}{\tau} \frac{\sigma_c}{\hat\sigma} + 1
 %\approx \frac{t_B}{\tau} \frac{\sigma_c}{\hat\sigma}
 \gg \frac{t_r(-\hat\sigma)}{\tau}
 \,,\label{eq:trhighstress}
\end{align}
and in this limit $t_r^\txt{eff}\approx t_r(+\hat\sigma)$.

\paragraph{Kinetic equation:} Taking into account both the remobilization time $t^\txt{eff}_r$ and the drag-limited transit time between network dislocations $t_{T0}=B\cL_i/(b\hat\sigma)$ (which currently neglects dislocation inertial effects), the mean dislocation velocity is $v=\cL_i/(t_r^\txt{eff}+t_{T0})$.

In Ref. \cite{Hunter:2015} it was argued that $v$ is underestimated by this relation.
The actual transit time will be less than $t_{T0}$ because the equation for $t_{T0}$ does not account for any distance the bowing dislocation has covered while pinned for time $t^\txt{eff}_r$.
This is especially true when $t^\txt{eff}_r(\hat\sigma)$ is close to $t_{T0}(\hat\sigma)$.
A correction factor $v\to\zeta v$ was developed in Ref. \cite{Hunter:2015} to address this inconsistency:
\begin{align}
 \zeta(\hat\sigma)\approx \left(1-\frac14\exp\left(-
\log^2_{10}\left[\frac{t_r^\txt{eff}(\hat\sigma)}{t_{T0}(\hat\sigma)}\right]\right)\right)^{-1} > 1
\,,
\end{align}
which typically yields at most a $10\%-15\%$ correction at $t_r(\hat\sigma)\sim t_{T0}(\hat\sigma)$.

The plastic strain rate, and thus the kinetic equation, finally follows from Orowan's relation \cite{Hunter:2015}:
\begin{align}
 \dot\vare &= \rho_m b\, v(\hat\sigma) = \frac{\zeta(\hat\sigma)\rho_m b}{\sqrt{\rho_i}t_r^\txt{eff}(\hat\sigma)+\frac{B(\hat\sigma)}{b\hat\sigma}}
 \,.\label{eq:kineticeqn}
\end{align}

We point out here that the kinetic equation (and thus the inverse kinetic equation described next) is dependent on both the mobile and immobile dislocation densities.
This model is meant to be combined with two evolution equations for the mobile and immobile dislocation densities.
While the kinetic equation could be coupled with any evolution equations for mobile and immobile dislocation density populations, model development for these equations is currently work in progress by co-authors of this work, \cite{Hunter:wip}.
Their formulation of these equations derives from theoretical development, stemming from dislocation theory, of terms accounting for specific dislocation storage, recovery, and generation mechanisms that have been observed experimentally in fcc metals.
In this work these dislocation densities can be varied to help provide understanding as to the impact on high rate material behavior, however they will not be dynamically updated during simulations presented in Section \ref{sec:results}.

\paragraph{Inverse kinetic equation:}
As mentioned previously, many continuum mechanics codes are velocity or strain rate driven, hence a formulation of the kinetic equation that gives stress as a function of strain rate is more useful.
However, since \eqnref{eq:kineticeqn} is a non-linear function of stress it cannot be inverted analytically in general.
It is possible to derive approximations for the inverse kinetic equation in certain limits.
Ref. \cite{Hunter:2015} considered the simple case of constant drag coefficient $B(\hat\sigma)=B_0$ and made use of the following additional approximations: In the high stress limit, $t_r^\txt{eff}$ can be replaced by \eqref{eq:trhighstress} and $\zeta\approx1$.
Unless $\frac{\hat\sigma}{\sigma_c}\gg1$, one can make the additional approximation $\frac{t_B}{\tau} \frac{\sigma_c}{\hat\sigma} + 1 \approx \frac{t_B}{\tau} \frac{\sigma_c}{\hat\sigma} $.
If we introduce new variables
\begin{align}
 x&\coleq \log_{10}\left(\dot\vare/\txt{s}^{-1}\right)
 \,, &
 y &\coleq \log_{10}\left(\hat\sigma/\sigma_c\right)
\end{align}
the high stress limit of the kinetic equation can be inverted to give
\begin{align}
 y&\approx m_+x+b_+
 \,,  \qquad\qquad
 m_+=1
 \,, \nn\\
 b_+ &= \log_{10}\left[\frac{\txt{s}^{-1}}{\rho_m b}\left(\sqrt{\rho_i}t_B+\frac{B}{b\sigma_c}\right)\right]
 =\log_{10}\left[\frac{\txt{s}^{-1}}{\rho_m b}\left(\frac{2\phi_c}{\pi^2}+1\right)\frac{B}{b\sigma_c}\right]
 \,,\label{eq:defbplusold}
\end{align}
where \eqnref{eq:estimates} was used in the last step.
% With $\phi_c=1/2$, this means that bow out time $t_B$ adds roughly 10\%  to the run time in the high stress regime.
%%%% not really run time: coincides with the run time only when sigma=sigma_c
% Notice that, $\sqrt{\rho_i}t_B=B/(\sqrt{\rho_i}\pi^2\cT)\ll B/(b\sigma_c)=B/(2\sqrt{\rho_i}\phi_c\cT)$, because $1/\pi^2\ll1/(2\phi_c)=1$.

In the case of low stresses, inversion of the kinetic equation proved more difficult.
Hence, additional complications arising in this regime required additional and more restrictive assumptions.
Specifically, these assumptions were:
\begin{enumerate}
 \item the up to $15\%$ correction from $\zeta$ was ignored for simplicity
 \item the drag coefficient was neglected in the low stress regime based on the assumption that $t_r\gg t_{T0}$ in this regime
 \item the remaining non-linear function $\log_{10}\left(t_r(y)\right)$ was approximated by a line giving rough agreement only in the regime $-1\lesssim y<0$:
 \begin{align*}
  \log_{10}\left(\tau/t_r\right)&\approx{y A^2\log_{10}e}
  \,.
 \end{align*}
\end{enumerate}
Hence, the low stress inverted kinetic equation can be written as:
\begin{align}
 y&\approx m_-x+b_-
 \,, &
 m_-&=\frac{1}{A^2\log_{10}e}
 \,, &
 b_-&=-m_-\log_{10}\left(\frac{\rho_m b}{\sqrt{\rho_i}\tau\txt{s}^{-1}}\right)
 \,. \label{eq:bminus_old}
\end{align}
We note that, due to these assumptions, the inverse kinetic equations given in Ref. \cite{Hunter:2015} cannot be used at very low stresses of $y\ll-1$.
In addition, the second assumption listed above fails to be true at high temperatures.
In Section \ref{sec:newmodel} below we eliminate this latter assumption and obtain an improved inverse kinetic equation.
Even with this improvement, it is recommended that the kinetic equation \eqref{eq:kineticeqn} be used when higher accuracy is required, especially in high temperature-low stress regimes.

The full inverse kinetic equation was then approximated by interpolating between the low and high stress regimes, using interpolation width $\Delta x\approx 0.75$:
\begin{align}
 y(x)&\approx \frac12\left[(m_+ + m_-)x + b_+ + b_-\right] + (m_+ - m_-)\frac{\Delta x}{2}\ln\left[2\cosh\left(\frac{x-x_c}{\Delta x}\right)\right]
 \,,\nn\\
 x_c&=\frac{b_- - b_+}{m_+ - m_-}
 \,. \label{eq:inversekinetic_paper}
\end{align}

\begin{table*}[h!t!b]
{\renewcommand{\arraystretch}{1.2}
\small
\centering
 \begin{tabular}{c|c|c|c|c|c|c|c|c}
 & $\rho_0$[g/ccm] & $G_0$[GPa] & $b_0$[\r{A}] & $T_{m}(\rho_0)$[K] & $\beta$ & $\gamma_1$[(g/ccm)$^{\frac13}$] & $\gamma_2$[(g/ccm)$^{q}$] & $q$ \\\hline\hline
 Al & 2.73 & 29.3 & 2.852 & 1277 & 0.18 & 0.84 & 45.4 & 3.5\\
 Cu & 9.02 & 52.4 & 2.55 & 1824 & 0.20 & 1.87 & 23100 & 4.7
\end{tabular}
\caption{%
Various material dependent model parameters used in the kinetic equations for Al and Cu.
Values were taken from Refs. \cite{Preston:1992,Burakovsky:2004}.
Variables $\rho_0$, $G_0$, $b_0$, and $T_{m0}$ denote density, shear modulus, and Burgers vector at zero Kelvin as well as the melting temperature at density $\rho_0$.
Density and shear modulus were taken from Ref. \cite{Burakovsky:2003}, $b_0$ and $T_{m}(\rho_0)$ were determined from \eqref{eq:shearmod} using room temperature data of Table \ref{tab:data_rt}.}
\label{tab:model-parameters}
}
\end{table*}

\paragraph{Density and temperature dependence:} Model parameters were assumed to be density and temperature independent in Ref. \cite{Hunter:2015}, while the Burgers vector and dislocation densities scale as:
\begin{align}
 b(\rho)&=b(\rho_0)\left(\frac{\rho_0}{\rho}\right)^{1/3}
 \,, &
 \rho_{i,m}(\rho) &= \rho_{i,m}(\rho_0)\left(\frac{\rho}{\rho_0}\right)^{2/3}
 \,. \label{eq:densityscaling}
\end{align}
A model for the density and temperature dependence of the shear modulus was derived in Refs. \cite{Preston:1992,Burakovsky:2003,Burakovsky:2004} and used in Ref. \cite{Hunter:2015}:
\begin{align}
 G(\rho,T) &= G(\rho,0)\left(1-\beta\frac{T}{T_m(\rho)}\right)
 \,,\nn\\
 T_m(\rho) &= T_m(\rho_0)\left(\frac{\rho}{\rho_0}\right)^{1/3}\exp\left\{6\gamma_1\left(\frac1{\rho_0^{1/3}}-\frac1{\rho^{1/3}}\right) + \frac{2\gamma_2}{q}\left(\frac1{\rho_0^{q}}-\frac1{\rho^{q}}\right)\right\}
 \,,\nn\\
%  G(\rho,0) &= G(\rho_0,0)\left(\frac{\rho}{\rho_0}\right)^{4/3}\left\{6\gamma_1\left(\frac1{\rho_0^{1/3}}-\frac1{\rho^{1/3}}\right) + \frac{2\gamma_2}{q}\left(\frac1{\rho_0^{q}}-\frac1{\rho^{q}}\right)\right\}
 G(\rho,0) &= G(\rho_0,0)\left(\frac{\rho\, T_m(\rho)}{\rho_0 \,T_m(\rho_0)}\right)
 \,, \label{eq:shearmod}
\end{align}
with material dependent model parameters $\beta$, $\gamma_1$, $\gamma_2$, and $q$ listed in Table \ref{tab:model-parameters} for fcc Al and Cu.
Room temperature data for these two metals are listed in Table \ref{tab:data_rt}.
$T_m(\rho)$ denotes the density dependent melt temperature.
Density and temperature dependence of the drag coefficient $B$ will be discussed in Section \ref{sec:dragcoeff} below, and all other quantities follow from their dependence on $b$, $\rho_{i,m}$, $G$, and $B$.

\begin{table*}[h!t!b]
{\renewcommand{\arraystretch}{1.2}
\small
\centering
 \begin{tabular}{c|c|c|c|c|c|c|c|c|c|c}
 & $a$[\r{A}] & $b$[\r{A}] & $\rho_\txt{rt}$[g/ccm] & $\ct$[m/s] & $\lambda$ & $G$ & $\ml$ & $\mm$ & $\mn$ & $T_m(\rho_\txt{rt})$[K] \\\hline\hline
 Al & 4.05 & 2.863 & 2.70 & 3109 & 58.1 & 26.1 & $-143$ & $-297$ & $-345$ & 933.47\\
 Cu & 3.61 & 2.556 & 8.96 & 2322 & 105.5 & 48.3 & $-160$ & $-620$ & $-1590$ & 1357.77
\end{tabular}
\caption{%
List of room temperature input data for aluminum and copper used in our calculations of the drag coefficient and the kinetic equations; all elastic constants are given in units of GPa.
The references we used to compile these data are: Ref.~\cite[Sec. 12]{CRCHandbook} (lattice parameters $a$, densities $\rh_\txt{rt}$, and melting temperatures at $\rh_\txt{rt}$),
Refs.~\cite[p.~10]{Hertzberg:2012} (effective Lam\'e constants of the polycrystal),
and \cite{Seeger:1960,Wasserbaech:1990} (Murnaghan constants).
The fcc Burgers vector lengths were determined by $b=a/\sqrt{2}$ and the transverse sound speeds were calculated from $\ct=\sqrt{G/\rho}$.
}
\label{tab:data_rt}
}
\end{table*}

\section{Dislocation drag in the isotropic limit}
\label{sec:dragcoeff}
%%%%%%%%%%%%%%%%%%%%%%%%%%%%%%%%%%%%%%%%

The viscous drag on objects moving through a fluid at sufficiently low velocities (low Reynolds numbers) is approximately proportional to the velocity;
the proportionality constant depends on the properties of the fluid and the geometry of the object.
Similarly, the drag force on a dislocation moving at velocities up to a few percent of $\ct$ (transverse sound speed) is proportional to the velocity, $F = B v$;
the drag coefficient, $B$, depends on the material properties and the character of the dislocation.
With increasing velocity the drag becomes highly nonlinear, ultimately diverging as $v\to\ct$.
The velocity dependence of the dislocation drag coefficient from phonon wind, the dominating contribution to drag at high stress and moderate to high temperatures, was discussed in a series of previous papers \cite{Blaschke:BpaperRpt,Blaschke:2018anis,Blaschke:2019fits,Blaschke:2019Bpap} which generalized earlier work of Alshits and collaborators reviewed in \cite{Alshits:1992}.
For a review of experimental work done on dislocation drag, see \cite{Nadgornyi:1988}.
MD simulation results on dislocation drag can be found e.g. in \cite{Rosakis:2001,Li:2002,Olmsted:2005,Jin:2008,Gilbert:2011,Ruestes:2015,Cho:2017,Chen:2017}.

\begin{figure}[ht]
	\centering
	\includegraphics[width=0.55\textwidth]{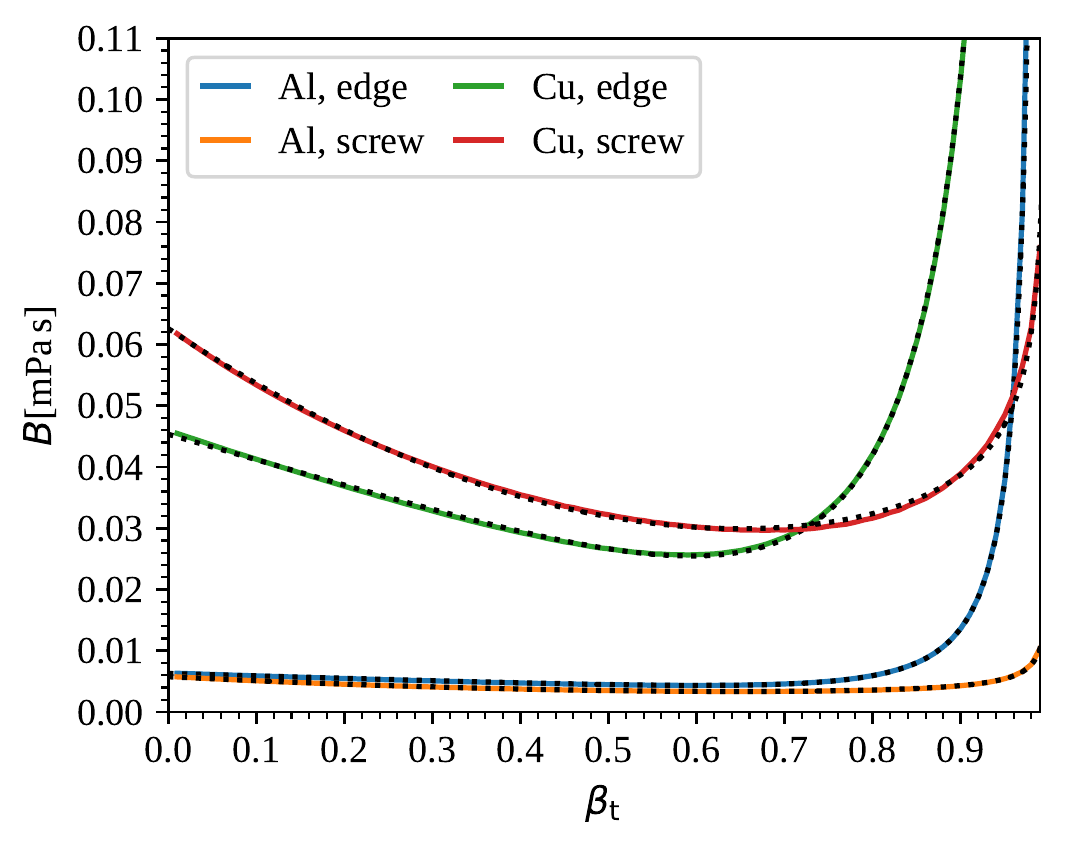}
	\caption{We compare the numerical solution to $B(v)$ for edge and screw dislocations (solid lines) to the fitting functions of \eqnref{eq:fittedcurves} (dashed lines) at the examples of Al and Cu in the range $0<\bt\le0.99$, see \cite{pydislocdyn}.}
	\label{fig:compareBfits}
\end{figure}

The energy dissipated per unit length of a moving dislocation is $D=Fv=Bv^2$ \cite{Alshits:1992,Blaschke:2019Bpap}.
In the first principles calculations of Refs. \cite{Alshits:1992,Blaschke:BpaperRpt,Blaschke:2018anis,Blaschke:2019fits,Blaschke:2019Bpap}, the dissipation $D$ was computed from the probability of scattering a phonon from state $q$ to state $q'$ per unit time and unit dislocation length.
The interaction Hamiltonian for this process was determined from a third-order expansion of the crystal potential in the continuum limit with respect to finite strain $\eta_{ij}$.
The phonon spectrum entering the calculation was approximated by an isotropic Debye spectrum.
The dislocation displacement gradient field, $u_{ij}$, can be calculated from stress-strain relations supplemented by an equation of motion even in the anisotropic case, provided the velocity is constant \cite{Bacon:1980}.
While Refs. \cite{Alshits:1992,Blaschke:BpaperRpt,Blaschke:2019Bpap} considered only the isotropic limit, Refs. \cite{Blaschke:2018anis,Blaschke:2019fits} calculated dislocation drag using the fully anisotropic dislocation field $u_{ij}$ and anisotropic second- and third-order elastic constants.
The latter references also included the interaction of dislocations with longitudinal phonons (in addition to transverse ones).

\begin{table*}[h!t!b]
{\renewcommand{\arraystretch}{1.2}
\small
\centering
 \begin{tabular}{c|c|c|c|c|c|c|c|c|c|c}
 & $C_0^{\text{e}}$ & $C_1^{\text{e}}$ & $C_2^{\text{e}}$ & $C_3^{\text{e}}$ & $C_4^{\text{e}}$ & $C_0^{\text{s}}$ & $C_1^{\text{s}}$ & $C_2^{\text{s}}$ & $C_3^{\text{s}}$ & $C_4^{\text{s}}$ \\\hline\hline
 Al & 6.3 & 4.1 & 1.5 & 0.0 & 1.2 & 5.7 & 6.9 & 4.3 & 0.3 & 1.4\\
 Cu & 45.3 & 41.9 & 72.4 & 136.4 & 3.6 & 62.6 & 96.6 & 65.5 & 0.0 & 7.8
\end{tabular}
\caption{%
Fitting parameters $C^{\txt{e}}_m$/$C^{\txt{s}}_m$ for edge/screw dislocations in Al and Cu at room temperature and ambient pressure in units of $\mu$Pa\,s.}
\label{tab:fitting-parameters}
}
\end{table*}

Here we restrict our discussion to the purely isotropic limit, consistent with the previous section, but do include longitudinal phonons (in contrast to Refs. \cite{Blaschke:BpaperRpt,Blaschke:2019Bpap}).
In order to numerically evaluate $B$ for aluminum and copper at ambient pressure and room temperature along the lines of those references, we employ the numerical implementation of the theory described in the previous paragraph, PyDislocDyn \cite{pydislocdyn}, 
which is written in Python and Fortran, is open source, and
developed by one of the authors.
The drag coefficient requires several integrals to be solved numerically, among them one two-dimensional integral over a non-trivial domain.
PyDislocDyn makes use of a trapezoidal method, with adaptive resolution where this is necessary;
for further details on the underlying theory we refer the interested reader to Refs. \cite{Blaschke:BpaperRpt,Blaschke:2019Bpap,Blaschke:2018anis}.

The results for $B$ can be accurately fit by the following functions, as shown in Figure \ref{fig:compareBfits}:
\begin{align}
 B_\txt{e}(\bt)&\approx C^\txt{e}_0 - C^\txt{e}_1\bt +C^\txt{e}_2\log\!\left(1-\bt^2\right) +C^\txt{e}_3\left({\left(1-\bt^2\right)^{-1/2}}-1\right) +C^\txt{e}_4\left({\left(1-\bt^2\right)^{-3/2}}-1\right)
 \,,\nn\\
 B_\txt{s}(\bt)&\approx C^\txt{s}_0 - C^\txt{s}_1\bt +C^\txt{s}_2\bt^2 +C^\txt{s}_3\log\!\left(1-\bt^2\right) +C^\txt{s}_4\left({\left(1-\bt^2\right)^{-1/2}}-1\right)
 \,, \label{eq:fittedcurves}
\end{align}
for edge ($B_\txt{e}$) and screw ($B_\txt{s}$) dislocations, where $\bt=v/\ct$ is the ratio of dislocation velocity and transverse sound speed.
Note the higher degrees of divergence compared to Refs.~\cite{Blaschke:BpaperRpt,Blaschke:2019Bpap}, where only the case of purely transverse phonons was considered and their asymptotic behavior was determined analytically for that special case.
A cancellation which reduced the degree of divergence for the transverse modes, does not occur for the mixed transverse-longitudinal modes leading to the increased degree of divergence of \eqref{eq:fittedcurves} when all phonon modes are taken into account.

Fitting parameters for Al and Cu were determined\footnote{
The results shown in Table \ref{tab:fitting-parameters} and Figure \ref{fig:compareBfits} can be reproduced with PyDislocDyn version 1.1.0 in its default settings by running:
\\\emph{python dragcoeff\_iso.py 'Al Cu'}.	
}
using PyDislocDyn \cite{pydislocdyn} and are listed in Table \ref{tab:fitting-parameters}.
Input parameters used in that calculation are listed in Table \ref{tab:data_rt}.
Since we do not distinguish between edge and screw dislocations in the (inverse) kinetic equation,
we will consider an average $B=\frac12\left(B_\txt{e} + B_\txt{s}\right)$ henceforth.

\paragraph{Temperature dependence:}
The drag coefficient depends on temperature via the phonon spectrum and the second (SOEC) and third-order (TOEC) elastic constants; the SOECs also affect the sound speeds.
At temperatures (on the order of or) higher than the Debye temperature, the isotropic phonon spectrum can be Taylor expanded, exhibiting a linear temperature dependence to leading order \cite{Alshits:1992,Blaschke:BpaperRpt}.
In fact, at constant density this is the dominating effect since the elastic constants are more sensitive to density and pressure than to temperature \cite{Preston:1992,Burakovsky:2003}.
Thus, as a rough approximation we may multiply $B$ by $T/300$ to incorporate its temperature dependence for $T\gtrsim300$ Kelvin.
Note, however, that this approximation cannot be expected to remain sufficiently accurate near the melting temperature.

\paragraph{Density dependence:}
In Ref. \cite{Alshits:1992,Blaschke:BpaperRpt} it was shown that the drag coefficient at low velocity, being dominated by transverse phonons, takes the form $B(v=0)\approx \frac{b^2q_\txt{BZ}^4}{\ct}f(\txt{SOECs,TOECs})$, where the Burgers vector is proportional to the lattice constant $a$ in a cubic lattice, the edge of the Brillouin zone scales as $q_\txt{BZ}\sim 1/a$, the transverse sound speed is determined from $\sqrt{G/\rho}$, and the function $f()$ represents a ratio of polynomials of SOEC and TOEC whose behavior under density changes is still unknown.
If all elastic constants scaled similarly with density, $f$ would hardly change.
Since we currently do not know how $f$ behaves under density changes, we neglect this effect for now.
The remaining variables then yield
\begin{align}
B(v=0,\rho,T)\approx B(v=0,\rho_\txt{rt},T)\frac{\rho}{\rho_\txt{rt}}\sqrt{\frac{\rho^{1/3} G(\rho_\txt{rt},T)}{\rho_\txt{rt}^{1/3} G(\rho,T)}}
\,, \label{eq:Bscaling}
\end{align}
where $\rho_\txt{rt}$ denotes the material density at room temperature and ambient pressure, and the shear modulus $G(\rho,T)$ was given in \eqnref{eq:shearmod}.
At high dislocation velocity, $B$ is dominated by $1/{(1-\bt^2)^{3/2}}$ with $\bt=v/\ct$, and in this regime the most important effect is the shift in limiting velocity $\ct(\rho,T)=\sqrt{G(\rho,T)/\rho}$.

\paragraph{Stress dependence:}
So far, we considered the dislocation drag coefficient as a function of velocity, but in real world applications dislocation velocity is not a quantity we can easily track.
Furthermore, dislocation velocity is the direct consequence of its driving (Peach-Koehler) force, and hence of the effective applied shear stress $\hat\sigma$, see  e.g. \cite{Luscher:2016}.
The governing equation allowing us to solve for $v(\hat\sigma)$ --- and subsequently derive $B(\hat\sigma)=B(v(\hat\sigma))$ at given density and temperature --- is:
\begin{align}
 b\,\hat\sigma = vB(v)
 \,,\label{eq:vofsigma}
\end{align}
which is consistent with \eqnref{eq:kineticeqn}, as in the short time \emph{between obstacles} $t^\txt{eff}_r=0$ (and $\zeta=1$), i.e. the drag equation above describes \emph{free running} dislocations.
In general, \eqref{eq:vofsigma} can only be solved for $v(\hat\sigma)$ numerically which is a viable path for the kinetic equation, but in order to generalize the inverse kinetic equation of the previous section we require a simple analytic form for $v(\hat\sigma)$.

\begin{figure}[ht]
 \centering
 \includegraphics[width=0.55\textwidth]{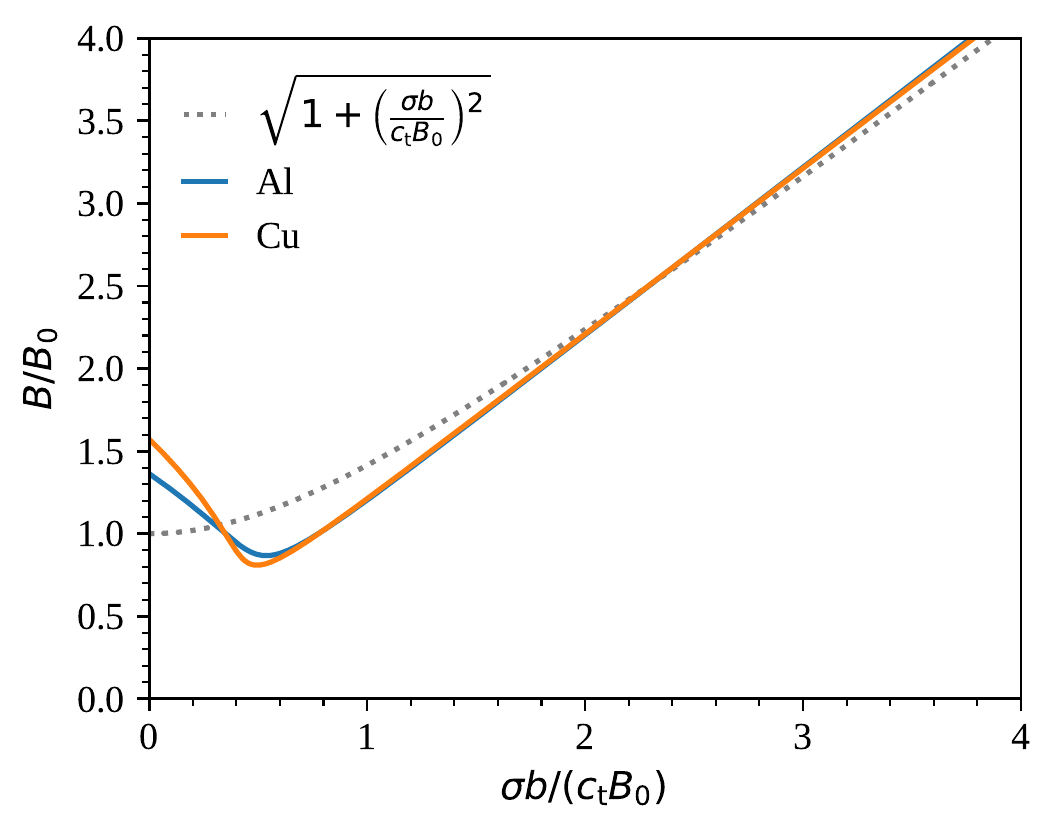}
 \caption{We compare the more accurate numerical solution to $B(\sigma)$ determined from \eqref{eq:fittedcurves} to the simple analytic function \eqref{eq:simpleBofsigma} (gray dashed curves) at the examples of Al and Cu,
 showing that \eqref{eq:simpleBofsigma} is indeed a reasonable approximation for the purpose of the (inverse) kinetic equation, provided (as we show in the next section below) that the kinetic equation is not very sensitive to the low stress regime of $B$.
 Parameters $B_0$ were determined from $(B(v=0) + 3\min(B(v)))/4$, leading to $B_0=4.4\mu$Pa\,s for Al and $B_0=34.3\mu$Pa\,s for Cu.
 }
 \label{fig:compareBsimple}
\end{figure}

Knowing that the (inverse) kinetic equation is dominated by drag effects especially in the high stress regime, we first examine the high stress limit of \eqref{eq:vofsigma}.
By high stress we mean stresses leading to dislocation velocities close to $\ct$, where $v$ cannot change much more with increasing stress.
In this regime, \eqref{eq:vofsigma} can be written as $B\approx b\hat\sigma/\ct$, exhibiting a linear stress dependence of $B$ in the asymptotic regime whose slope is $b/\ct$.
Since the stress regime leading to velocities much less than $\ct$ is fairly narrow compared to the broad range of high stresses we are interested in, the global behavior of $B$ is well approximated by the asymptotic terms, i.e. $B(v)\approx B_0/({1-\bt^2})^m$ with $m=1/2$ for screw and $m=3/2$ for edge dislocations, and
with only one fitting parameter $B_0$, to be chosen as an average value in the low to intermediate velocity regime of \eqref{eq:fittedcurves}.
With exponent $m=1/2$, one can analytically solve \eqnref{eq:vofsigma} to yield
\begin{align}
 B(\hat\sigma)&=\sqrt{B_0^2 + \left(\frac{b\,\hat\sigma}{\ct}\right)^2}
 \label{eq:simpleBofsigma}
 \,,
\end{align}
and in fact this functional form is in good agreement with $B$ computed as an average over screw and edge dislocations.
Figure \ref{fig:compareBsimple} compares the simple analytic form of \eqnref{eq:simpleBofsigma} to the numerically determined exact $B(\hat\sigma)=\left(B_\txt{e}(\hat\sigma)+B_\txt{s}(\hat\sigma)\right)/2$ for Al and Cu.
The deviations in the low stress regime are likely within the model uncertainties which are expected to be fairly large due to the omission of dislocation core and anisotropic effects.
At this point, it is therefore also unclear if the initial drop in magnitude of $B$ at low stresses (which is due to an interplay of energy conservation and integral ranges with the edge of the Brillouin zone \cite{Blaschke:BpaperRpt,Blaschke:2019Bpap}) is indeed a real effect or merely an artifact of a simplified model.
Furthermore, as we will see in the next section, the kinetic equations are not very sensitive to $B$ in the low to intermediate stress regimes.
In the high stress regime, \eqnref{eq:simpleBofsigma} tends to $\lim_{\hat\sigma\to\infty}B(\hat\sigma)\approx b\hat\sigma/\ct$, as expected.
Hence, we conclude that the approximation \eqref{eq:simpleBofsigma} is sufficient for our present purpose.

The dependence of $B$ on temperature and density, as discussed above, straightforwardly leads to
\begin{align}
 B(\hat\sigma,\rho,T)&\approx B_0(\rho,T)/\sqrt{1-\bt(\hat\sigma,\rho,T)^2}
 \,, \nn\\
 B_0(\rho,T) &= B_0(\rho_\txt{rt},300)\frac{T\rho}{300\rho_\txt{rt}}\sqrt{\frac{\rho^{1/3} G(\rho_\txt{rt},300)}{\rho_\txt{rt}^{1/3} G(\rho,T)}}
 \,, &
 \bt(\rho,T) &= \bt(\rho_\txt{rt},300)\frac{\ct(\rho_\txt{rt},300)}{\ct(\rho,T)}
 \,,\label{eq:simpleBofvTrho}
\end{align}
which yields
\begin{align}
 B(\hat\sigma,\rho,T)&=\sqrt{B_0(\rho,T)^2 + \left(\frac{b(\rho)\,\hat\sigma}{\ct(\rho,T)}\right)^2}
 \label{eq:simpleBofsigmaTrho}
 \,.
\end{align}
The only model parameter which needs to be determined from \eqref{eq:fittedcurves} is $B_0(\rho_\txt{rt},300)$.

\section{Generalization of the model}
\label{sec:newmodel}
%%%%%%%%%%%%%%%%%%%%%%%%%%%%%%%%%%%%%%%%

\begin{figure}[ht]
 \centering
 \includegraphics[width=0.5\textwidth]{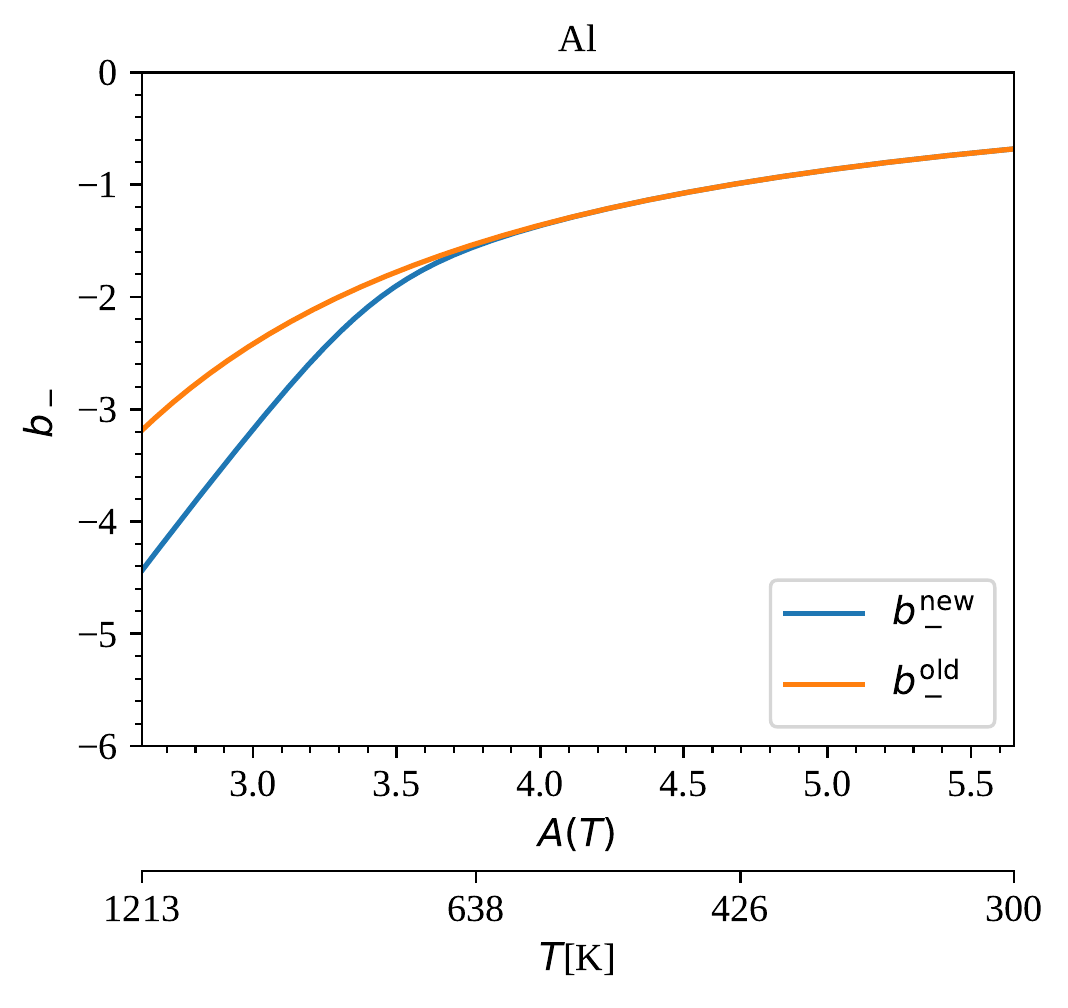}%
 \includegraphics[width=0.5\textwidth]{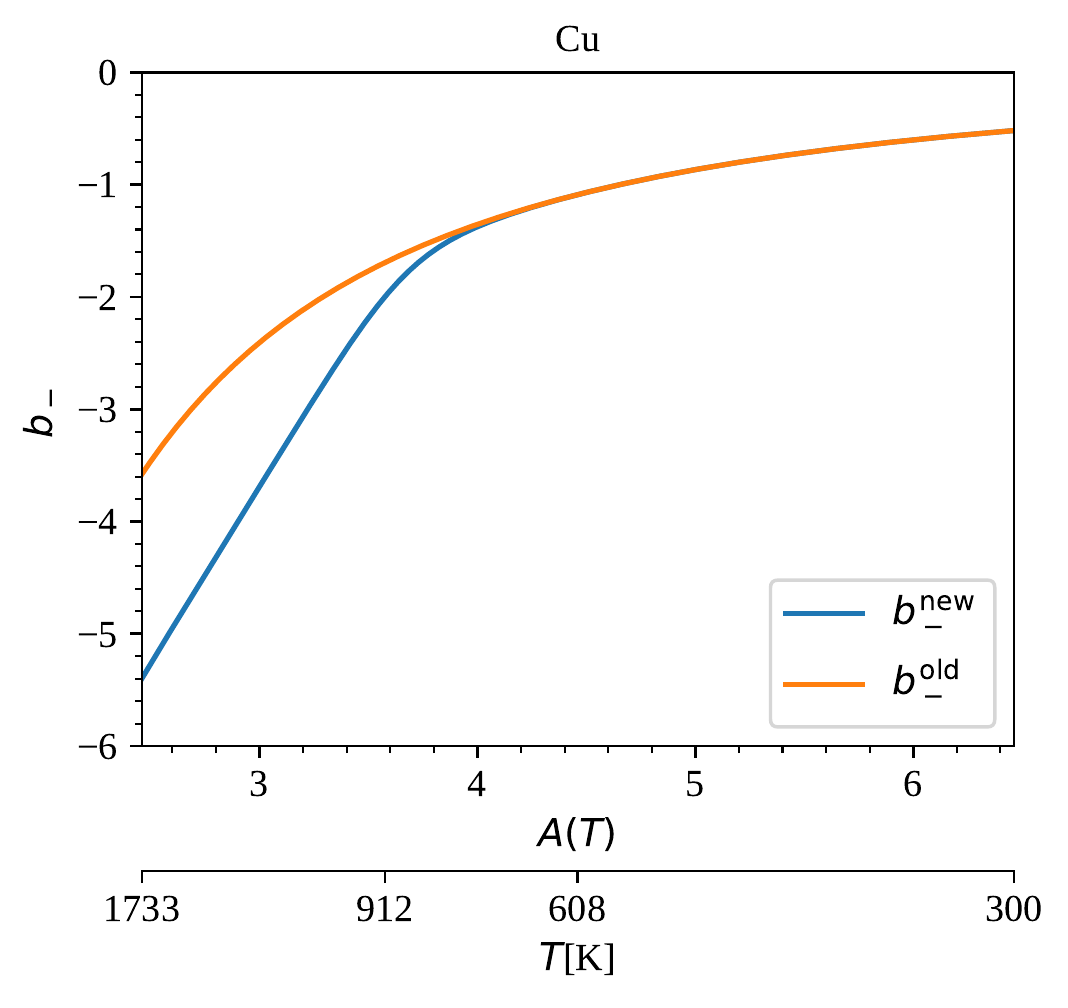}
 \caption{We compare $b_-^\text{old}$ as defined in \eqref{eq:bminus_old} to $b_-^\text{new}$ as defined in \eqref{eq:bminusnew}.
 The two $x$ axes show temperature as well as $A$ calculated for that temperature.
 The material and dislocation densities were taken to be $\rho=\rho_0$ and $\rho_i=10^{12}$m$^{-2}=\rho_m$.}
 \label{fig:bminus}
\end{figure}

\begin{figure}[!h!t]
 \centering
 \includegraphics[width=0.5\textwidth]{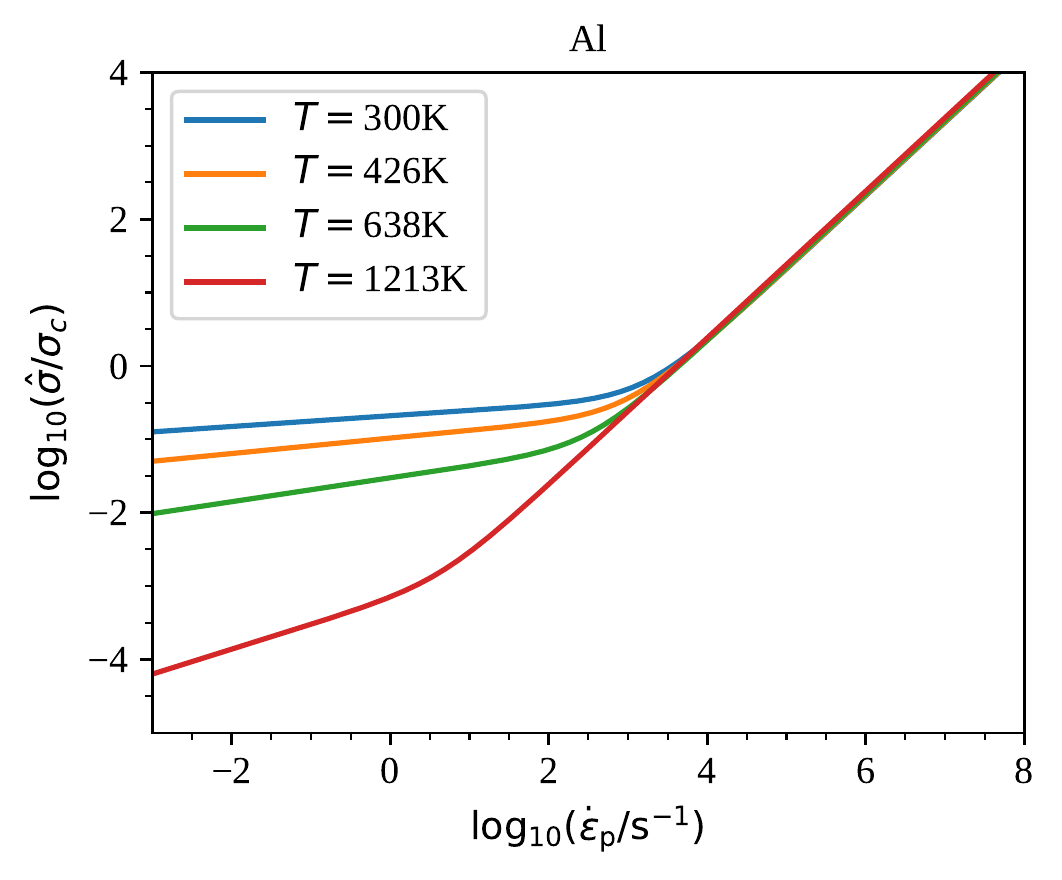}%
 \includegraphics[width=0.5\textwidth]{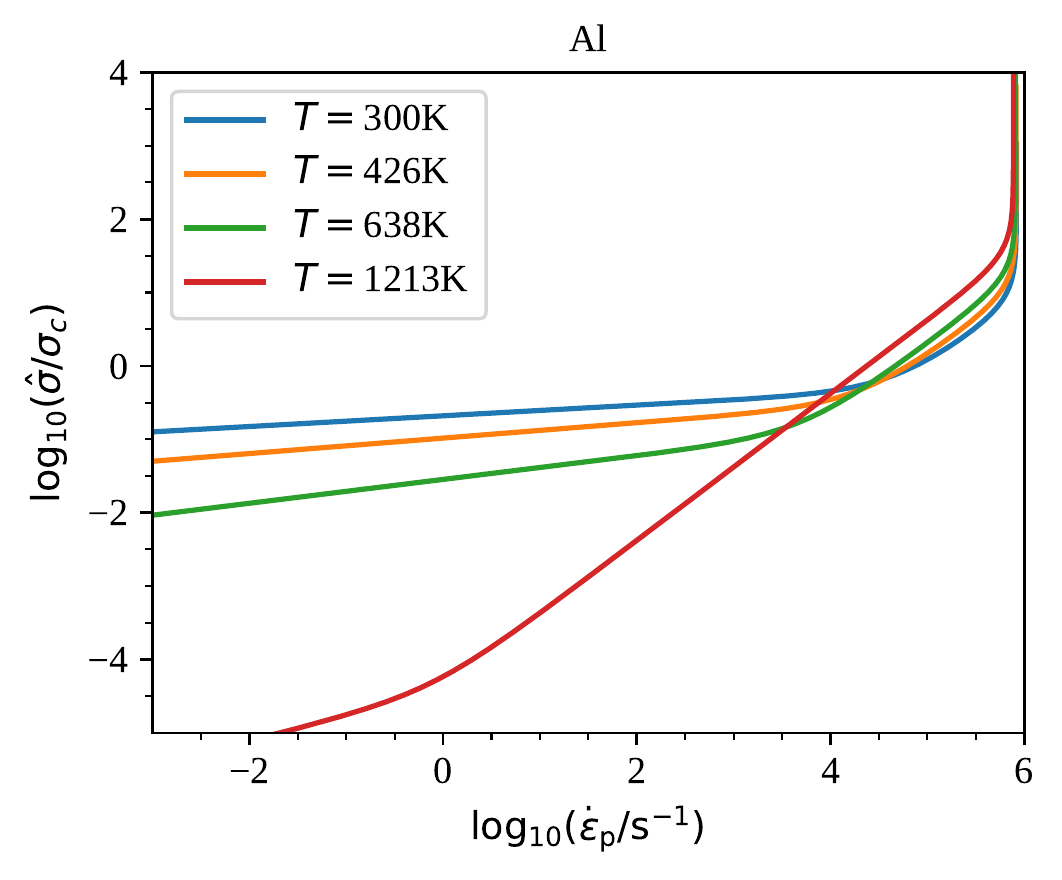}
 \includegraphics[width=0.5\textwidth]{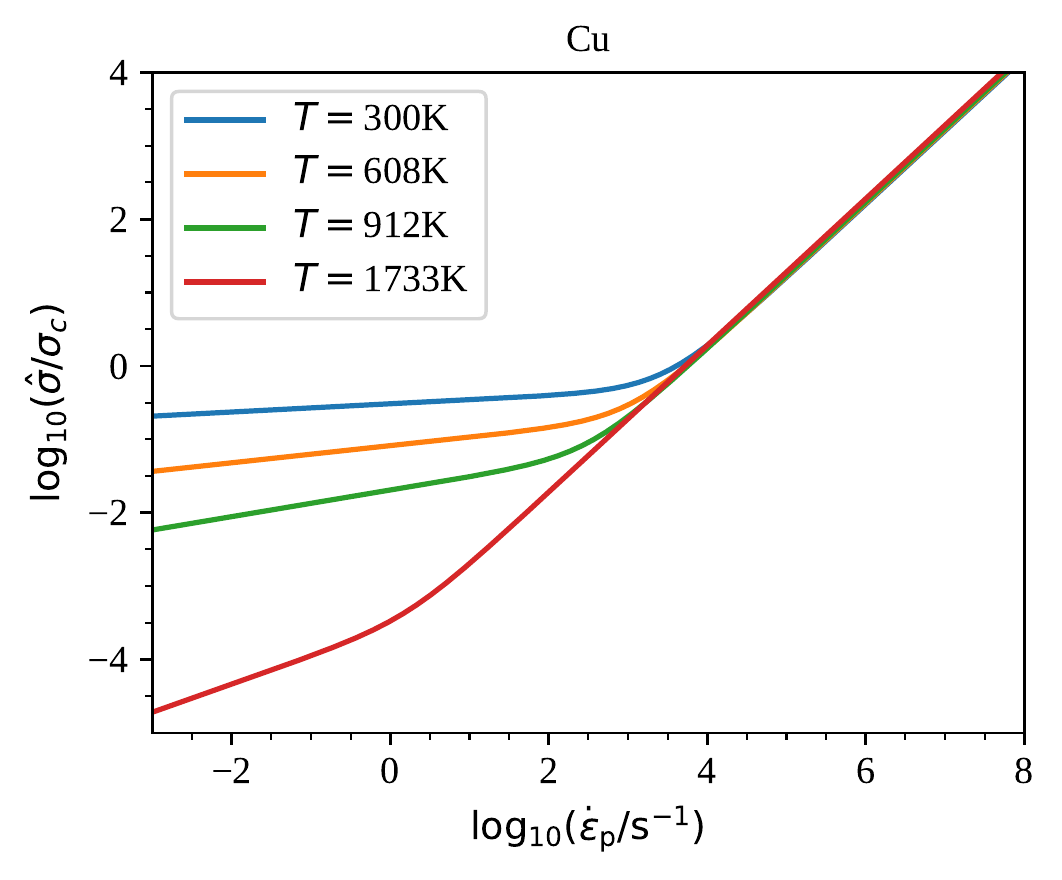}%
 \includegraphics[width=0.5\textwidth]{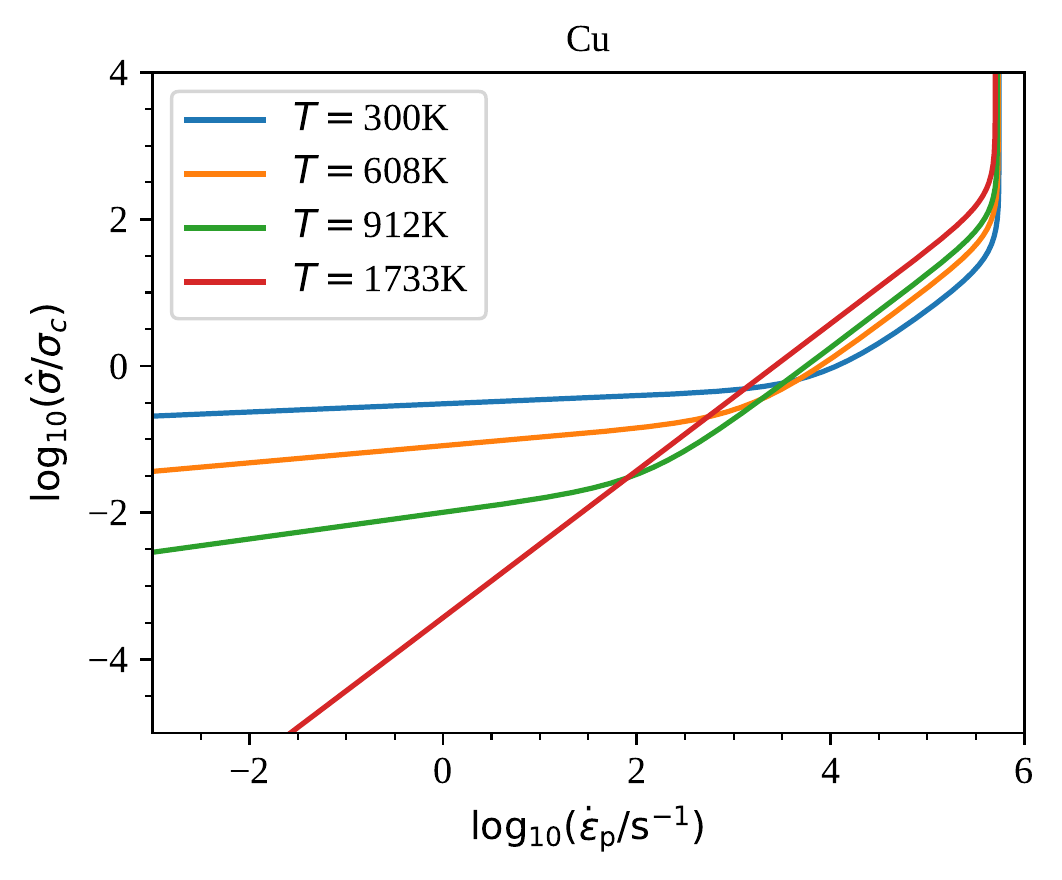}
 \caption{Comparison of the temperature dependence of $\hat\sigma/\sigma_c$ within the approximate inverse kinetic equation for aluminum and copper:
 The left column shows the approximation using the originally developed inverse kinetic equation (Eqns. \eqref{eq:defbplusold}, \eqref{eq:bminus_old}) and constant drag $B=0.1$mPas, as was considered in the earlier work \cite{Hunter:2015}, whereas in the right column the refined $b_\pm$ of \eqref{eq:bplusnew} (which includes $B(\hat\sigma,\rho,T)$) and \eqref{eq:bminusnew} were taken into account.
 %  \\
 Three differences are clearly visible: the maximum strain rate is due to $B$ diverging at a maximum dislocation velocity, the ``splitting'' of the curves in the intermediate region is due to the temperature dependence of $B$, and the refined $b_-$ of \eqref{eq:bminusnew} leads to a smaller $x_c$ (i.e. the position where the curves ``bend'' in the low strain rate regime) at high temperature.
 The material and dislocation densities were taken to be $\rho=\rho_0$ and $\rho_i=10^{12}$m$^{-2}=\rho_m$.}
 \label{fig:inversekinetic_Tdep}
\end{figure}

\begin{figure}[!h!t]
 \centering
 \includegraphics[width=0.5\textwidth]{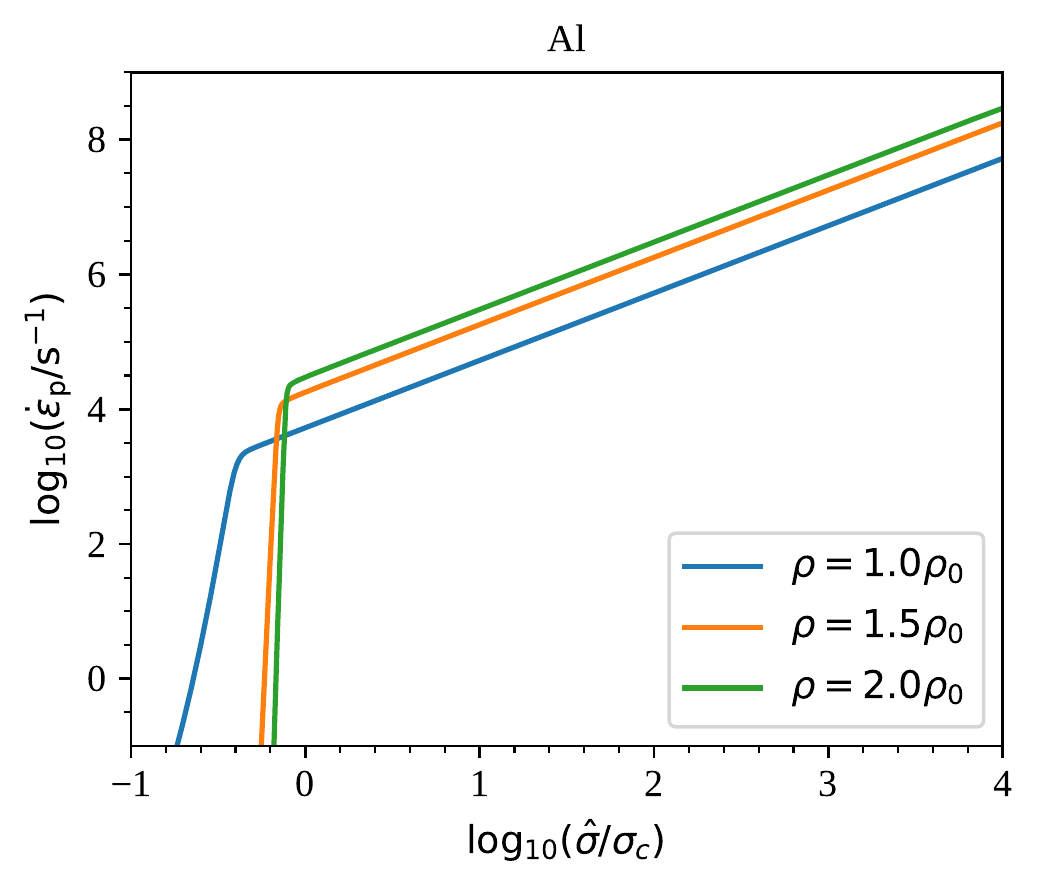}%
 \includegraphics[width=0.5\textwidth]{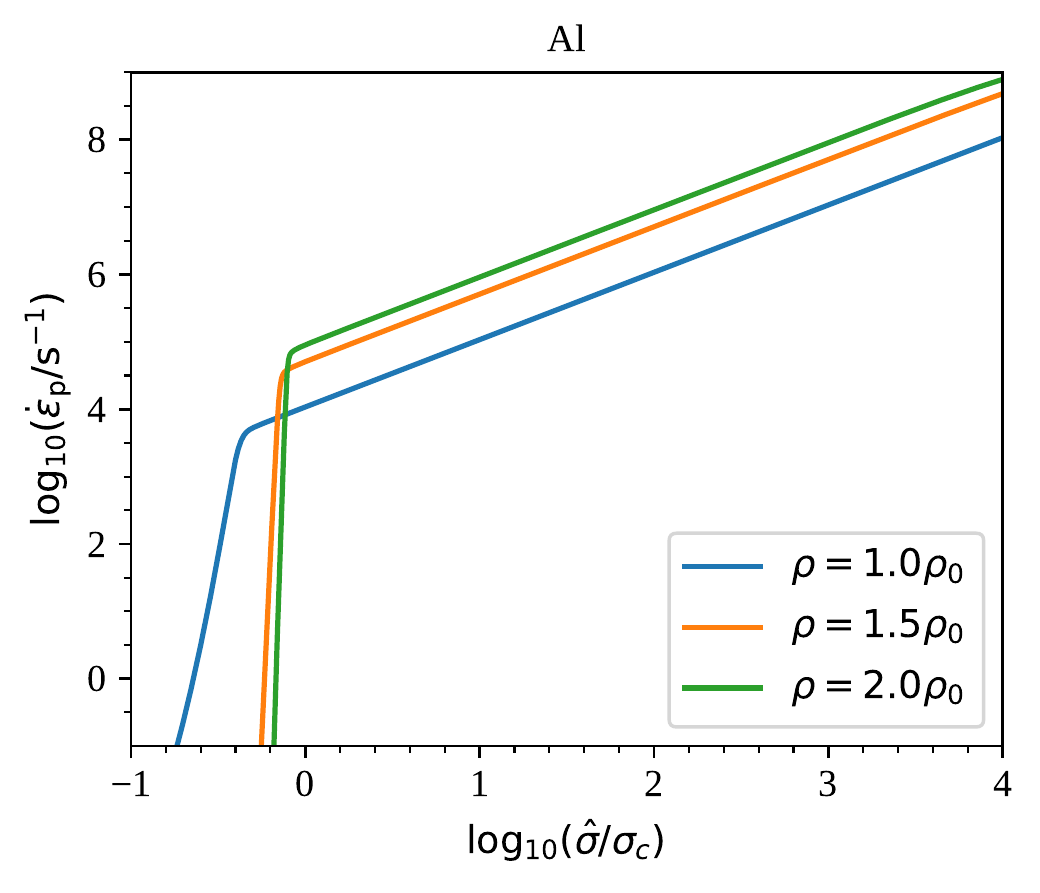}
 \includegraphics[width=0.5\textwidth]{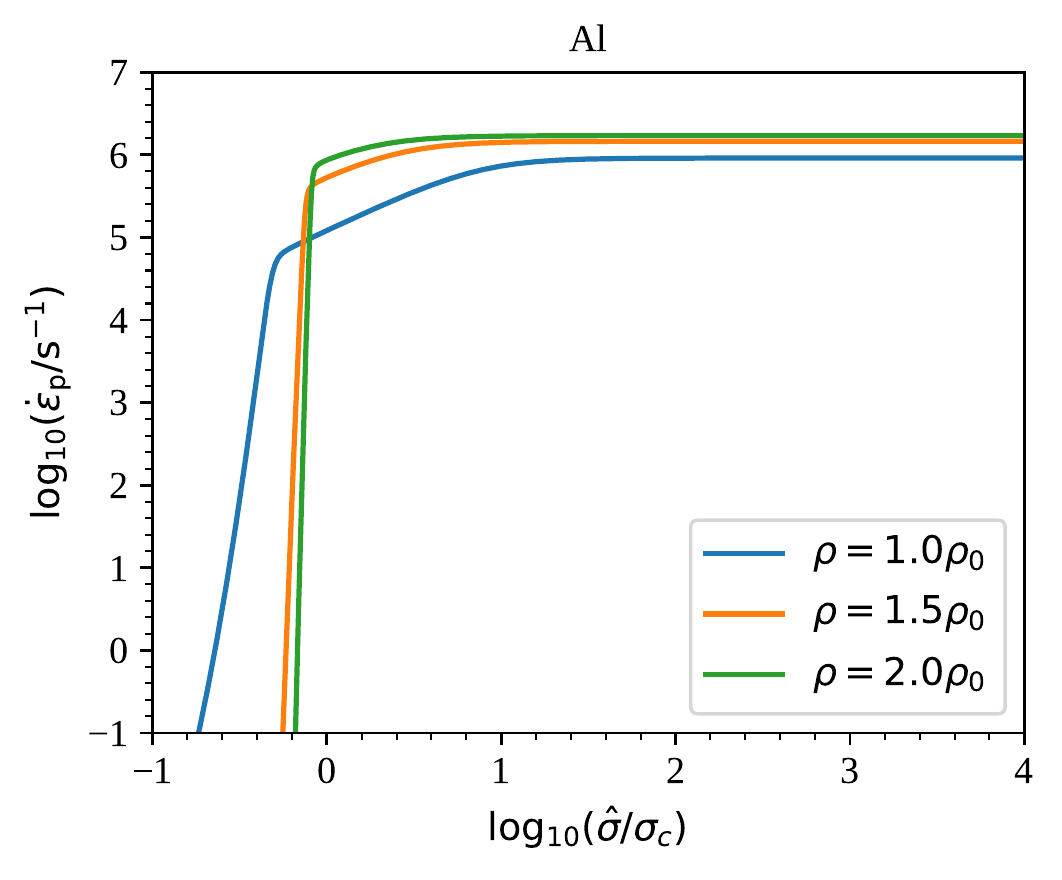}%
 \includegraphics[width=0.5\textwidth]{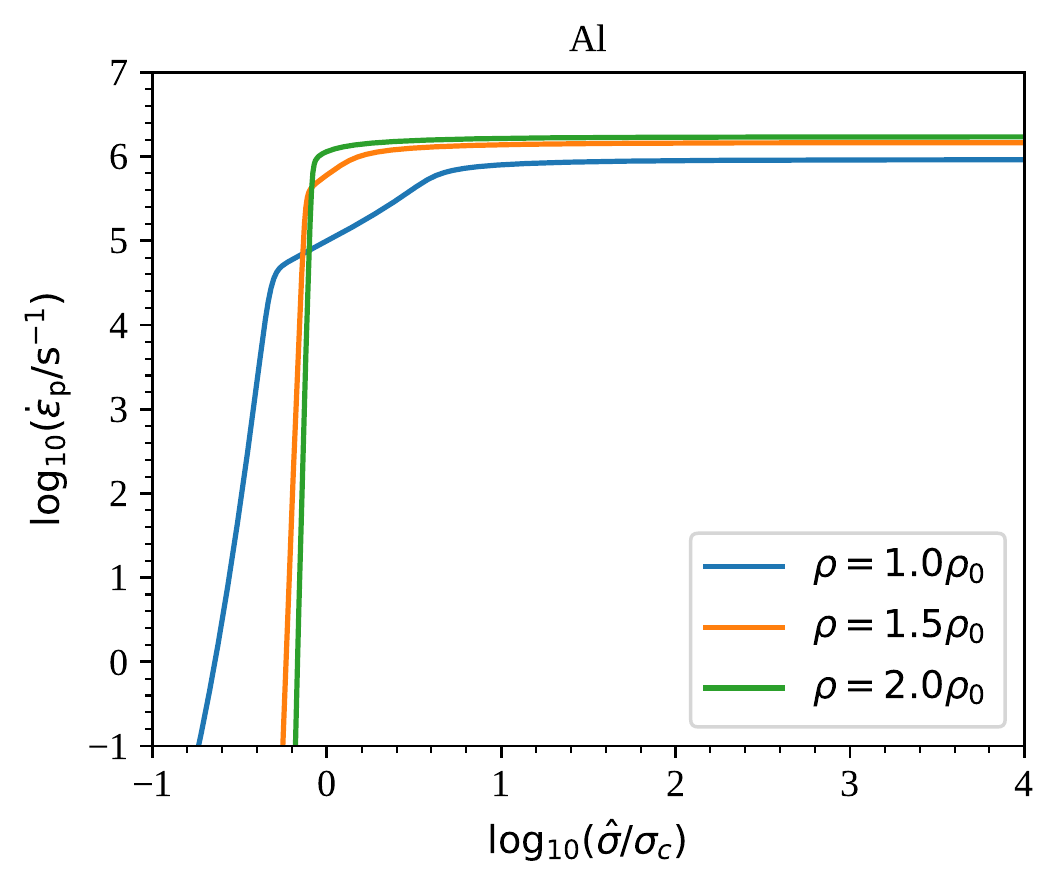}
 \caption{Material density dependence of $\dot\vare$ as given by the kinetic equation for aluminum for the case of a) constant drag coefficient $B=0.1$mPa\,s as used within the model of Ref.~\cite{Hunter:2015} (top left),
 b) drag coefficient $B(\hat\sigma=0,\rho,T)$ with $B_0=0.05$mPa\,s ($B_0$ at room temperature and ambient density) (top right),
 c) drag coefficient $B(\hat\sigma,\rho,T)$ in its simple functional form as given by \eqref{eq:simpleBofsigmaTrho} (bottom left),
 d) drag coefficient $B(\hat\sigma,\rho,T)$ in its full (numerically determined) form derived using \eqref{eq:fittedcurves} and \eqref{eq:vofsigma} (bottom right).
 Other parameters used in these figures are: $T=300$K and $\rho_i=10^{12}$m$^{-2}=\rho_m$.}
 \label{fig:kinetic_Al_rho0}
\end{figure}

\begin{figure}[!h!t]
 \centering
 \includegraphics[width=0.5\textwidth]{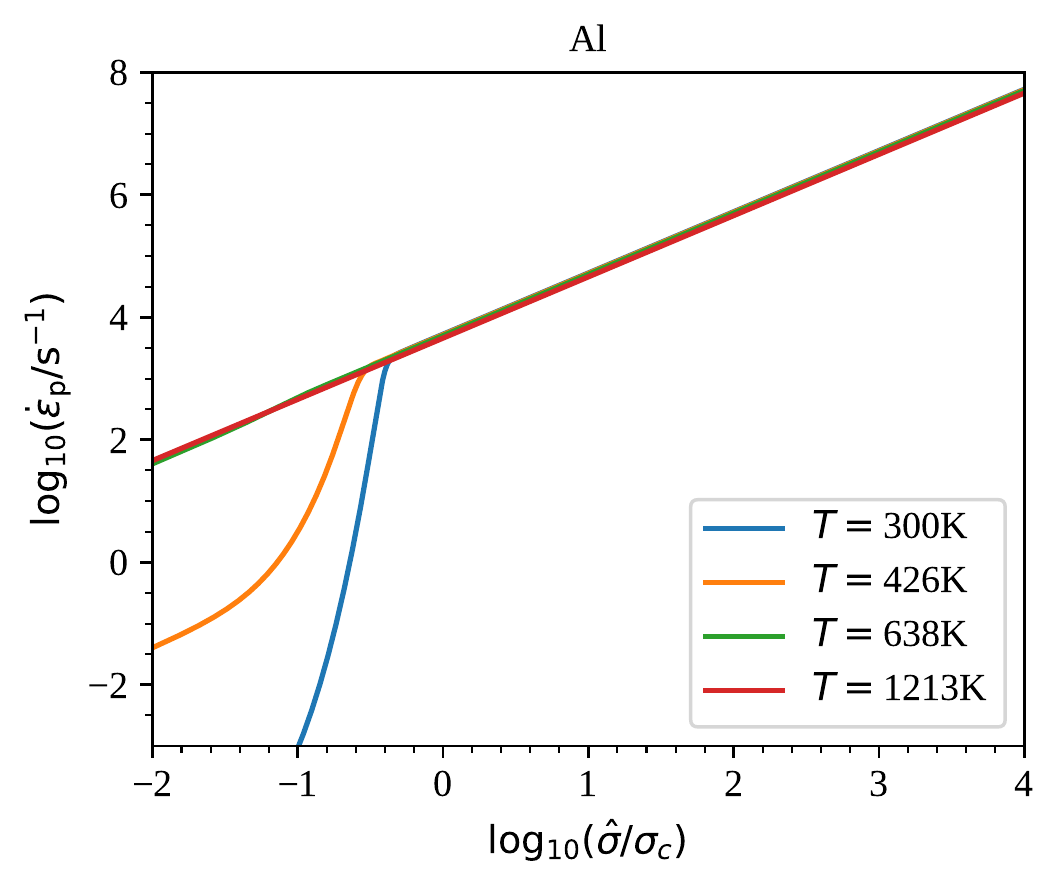}%
 \includegraphics[width=0.5\textwidth]{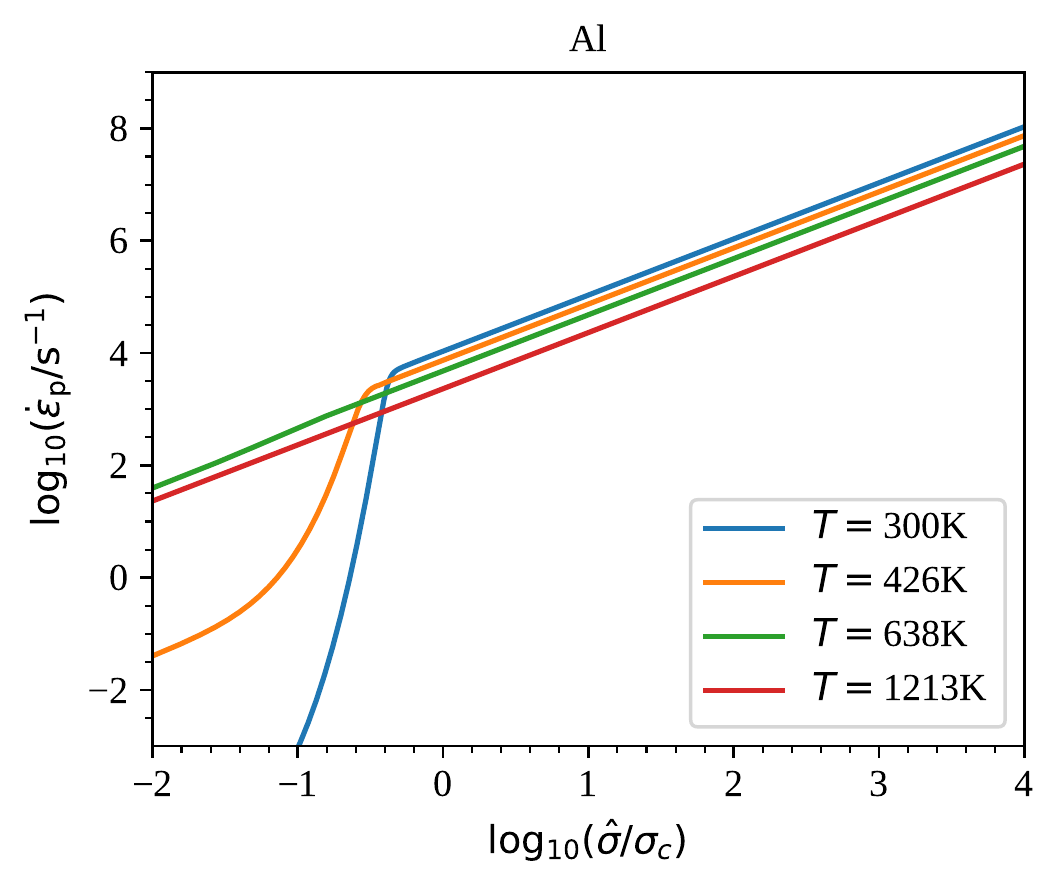}
 \includegraphics[width=0.5\textwidth]{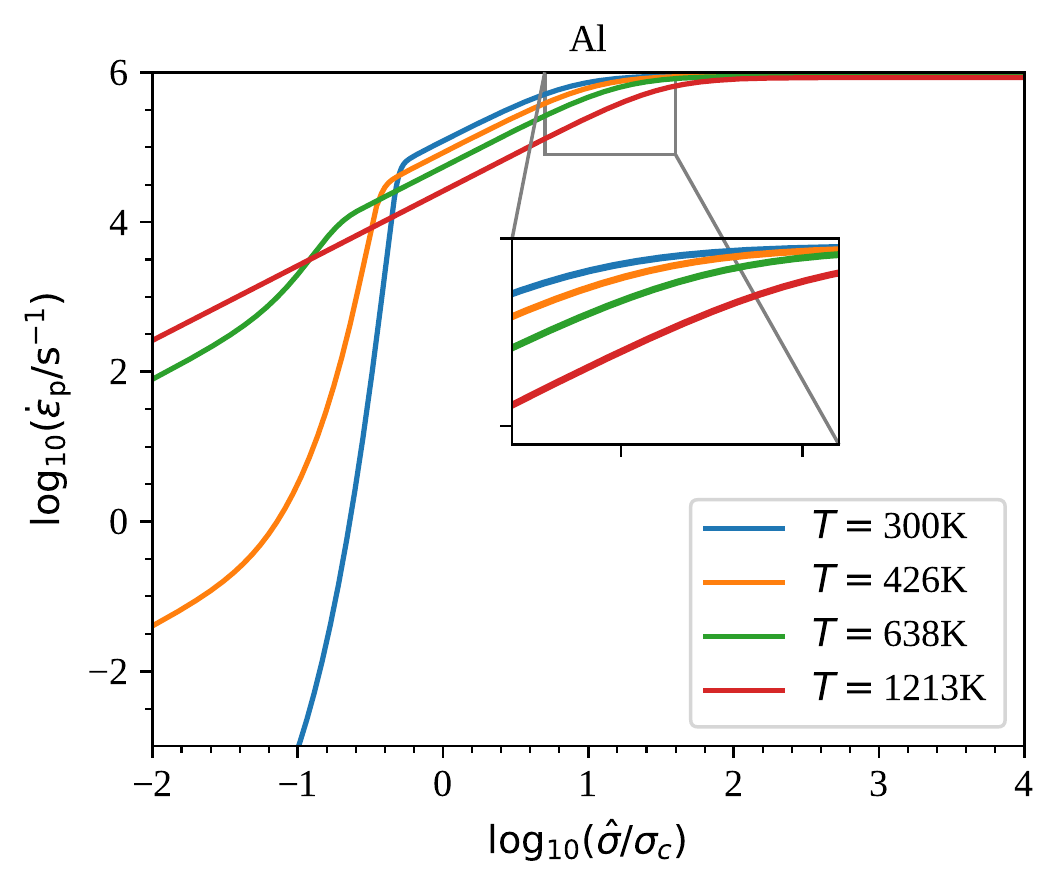}%
 \includegraphics[width=0.5\textwidth]{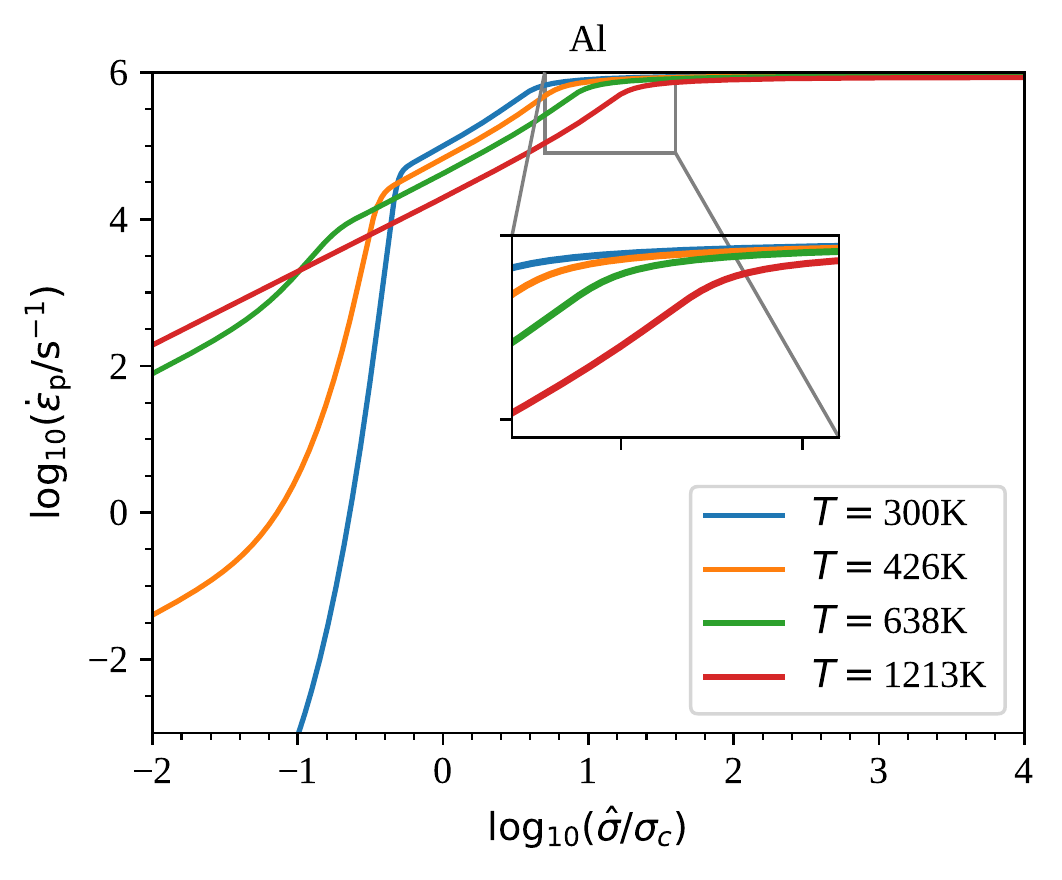}
 \caption{Temperature dependence of $\dot\vare$ as given by the kinetic equation for aluminum for the case of a) constant drag coefficient $B=0.1$mPa\,s as used within the model of Ref.~\cite{Hunter:2015} (top left),
 b) drag coefficient $B(\hat\sigma=0,\rho,T)$ with $B_0=0.05$mPa\,s ($B_0$ at room temperature and ambient density) (top right),
 c) drag coefficient $B(\hat\sigma,\rho,T)$ in its simple functional form as given by \eqref{eq:simpleBofsigmaTrho} (bottom left),
 d) drag coefficient $B(\hat\sigma,\rho,T)$ in its full (numerically determined) form derived using \eqref{eq:fittedcurves} and \eqref{eq:vofsigma} (bottom right).
 The two inset plots in  the bottom row highlight the differences between the latter two in the regime close to $\log_{10}\left(\hat\sigma/\sigma_c\right)\sim1$, which are otherwise difficult to see.
 Other parameters used in these figures are: $\rho=\rho_0$ and $\rho_i=10^{12}$m$^{-2}=\rho_m$.}
 \label{fig:kinetic_Al_T}
\end{figure}

\begin{figure}[!h!t]
 \centering
 \includegraphics[width=0.5\textwidth]{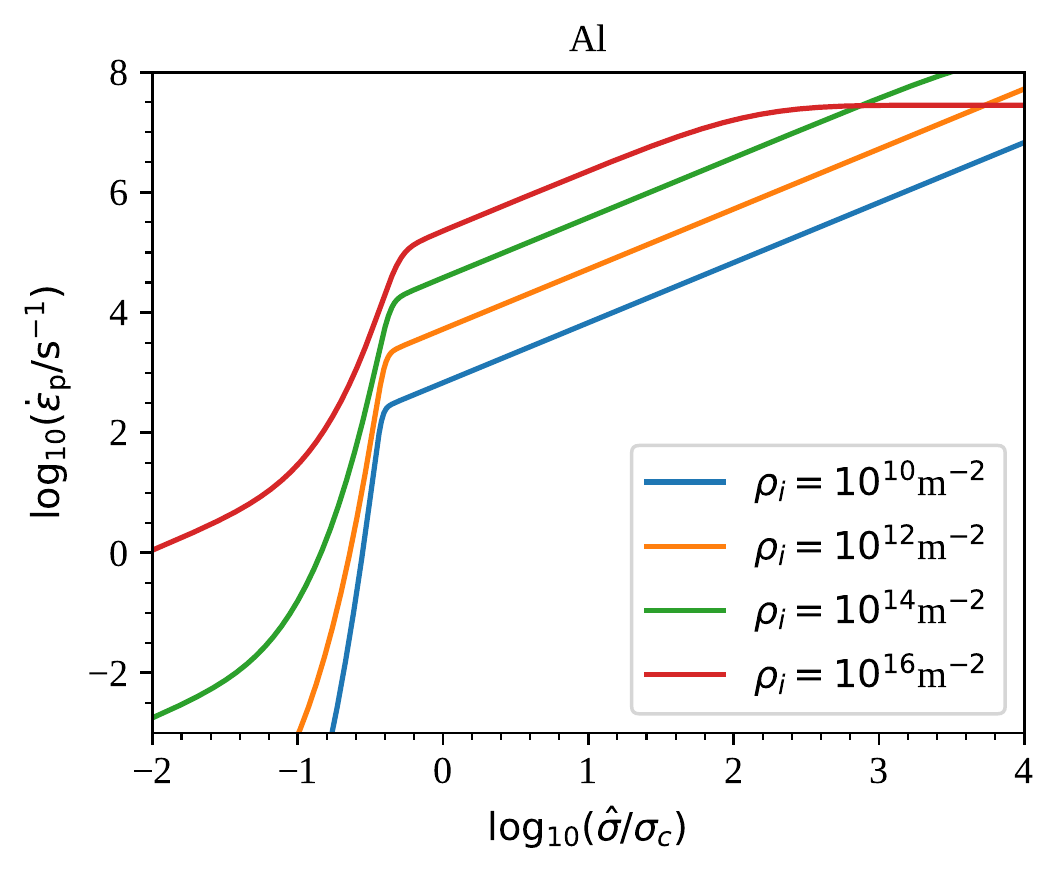}%
 \includegraphics[width=0.5\textwidth]{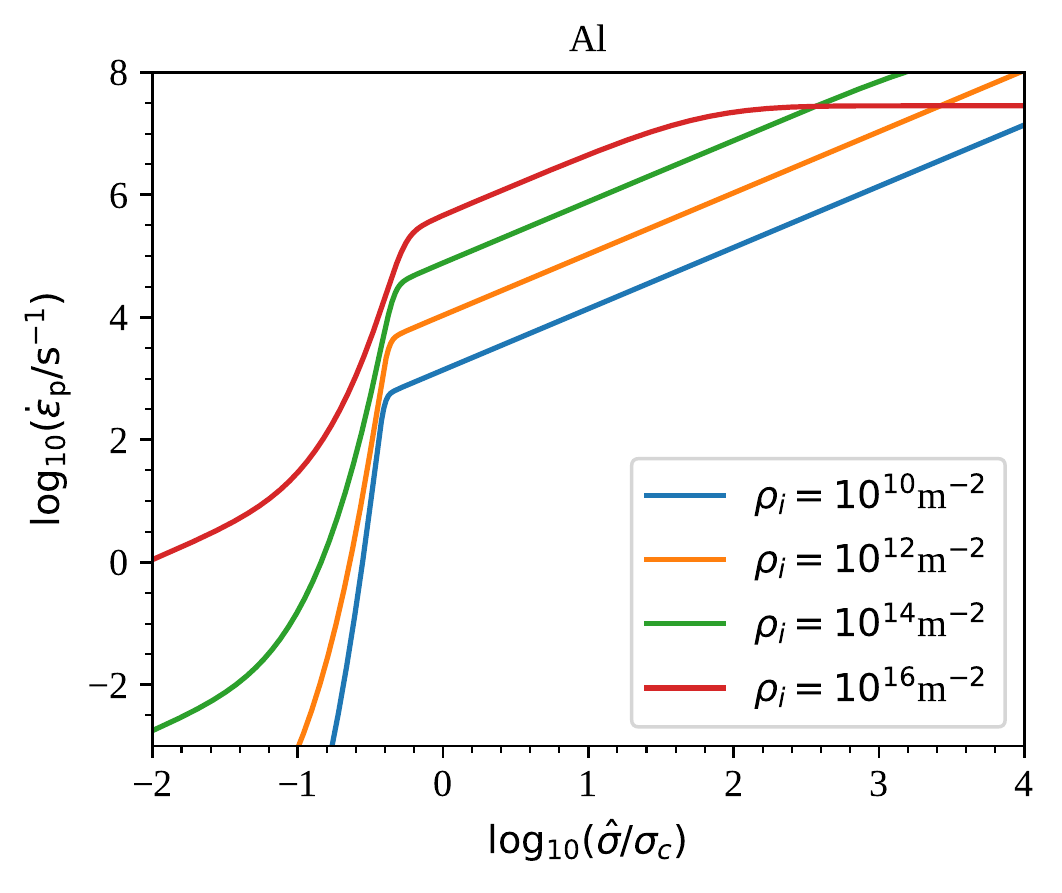}
 \includegraphics[width=0.5\textwidth]{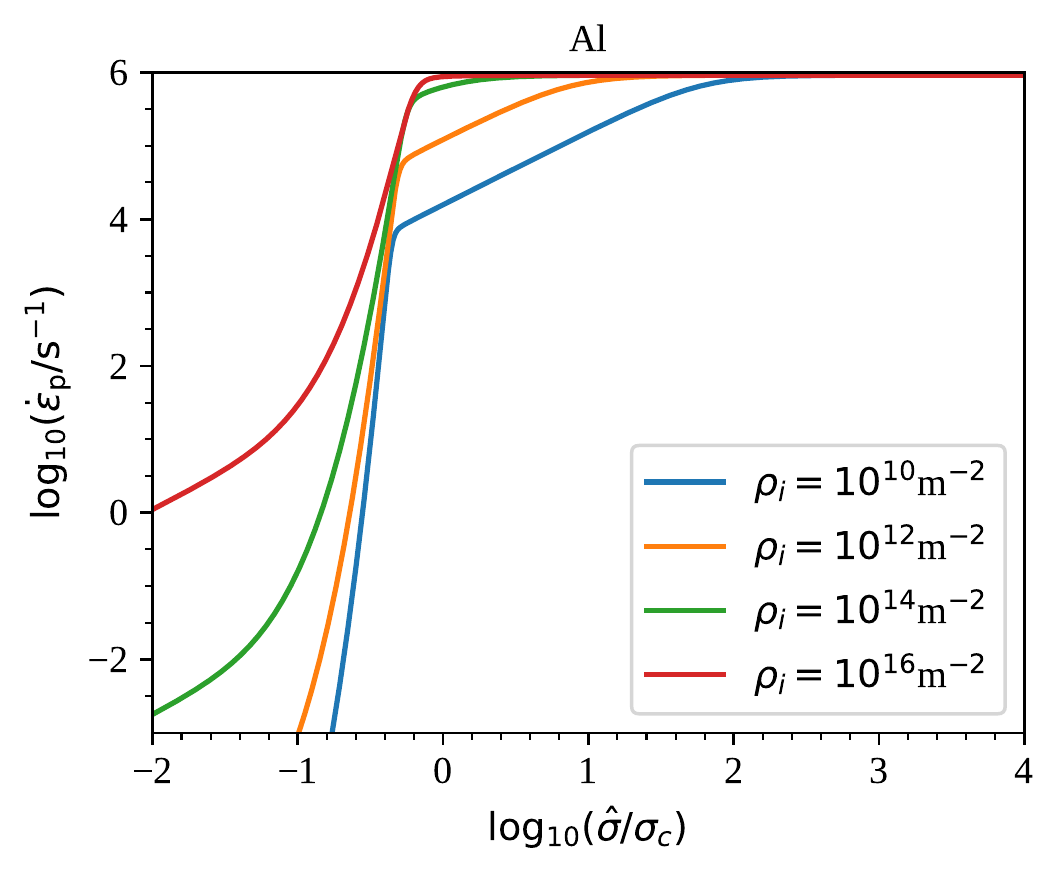}%
 \includegraphics[width=0.5\textwidth]{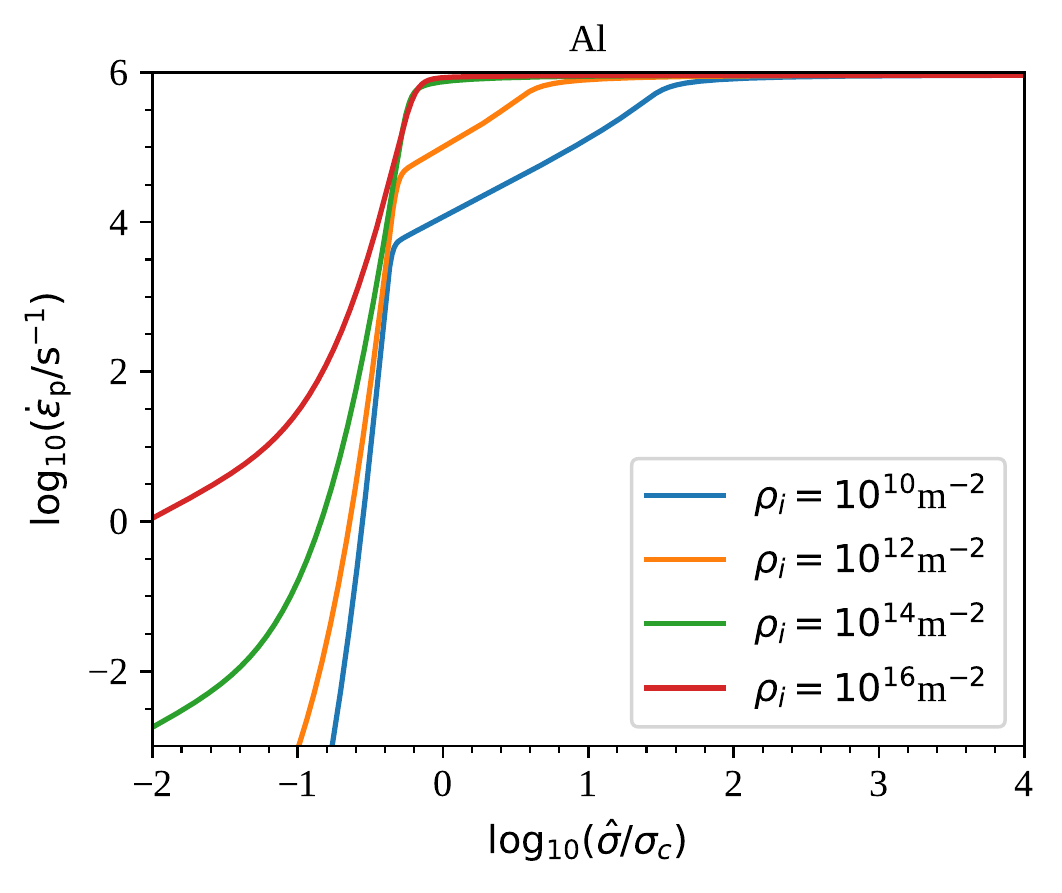}
 \caption{Immobile dislocation density dependence of $\dot\vare$ as given by the kinetic equation for aluminum for the case of a) constant drag coefficient $B=0.1$mPa\,s as used within the model of Ref.~\cite{Hunter:2015} (top left),
 b) drag coefficient $B(\hat\sigma=0,\rho,T)$ with $B_0=0.05$mPa\,s ($B_0$ at room temperature and ambient density) (top right),
 c) drag coefficient $B(\hat\sigma,\rho,T)$ in its simple functional form as given by \eqref{eq:simpleBofsigmaTrho} (bottom left),
 d) drag coefficient $B(\hat\sigma,\rho,T)$ in its full (numerically determined) form derived using \eqref{eq:fittedcurves} and \eqref{eq:vofsigma} (bottom right).
 Other parameters used in these figures are: $T=300$K, $\rho=\rho_0$ and $\rho_m=10^{12}$m$^{-2}$.}
 \label{fig:kinetic_Al_rhoi}
\end{figure}

\begin{figure}[!h!t]
 \centering
 \includegraphics[width=0.5\textwidth]{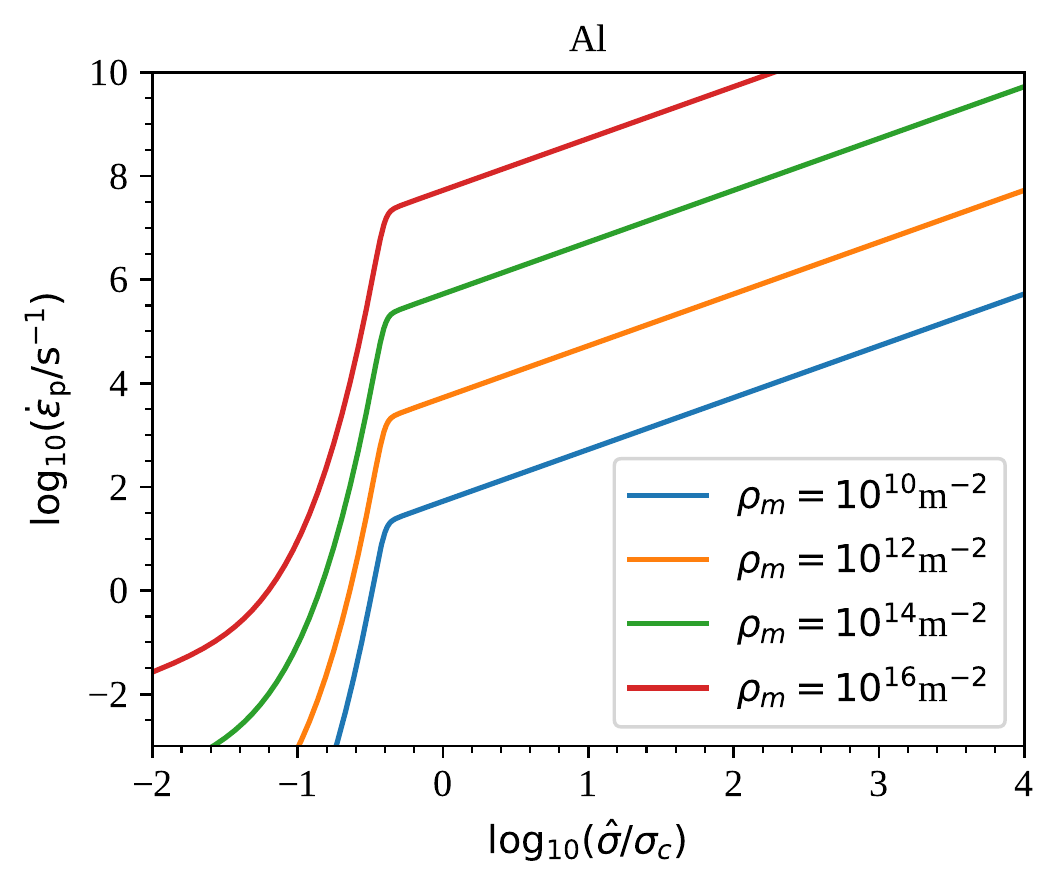}%
 \includegraphics[width=0.5\textwidth]{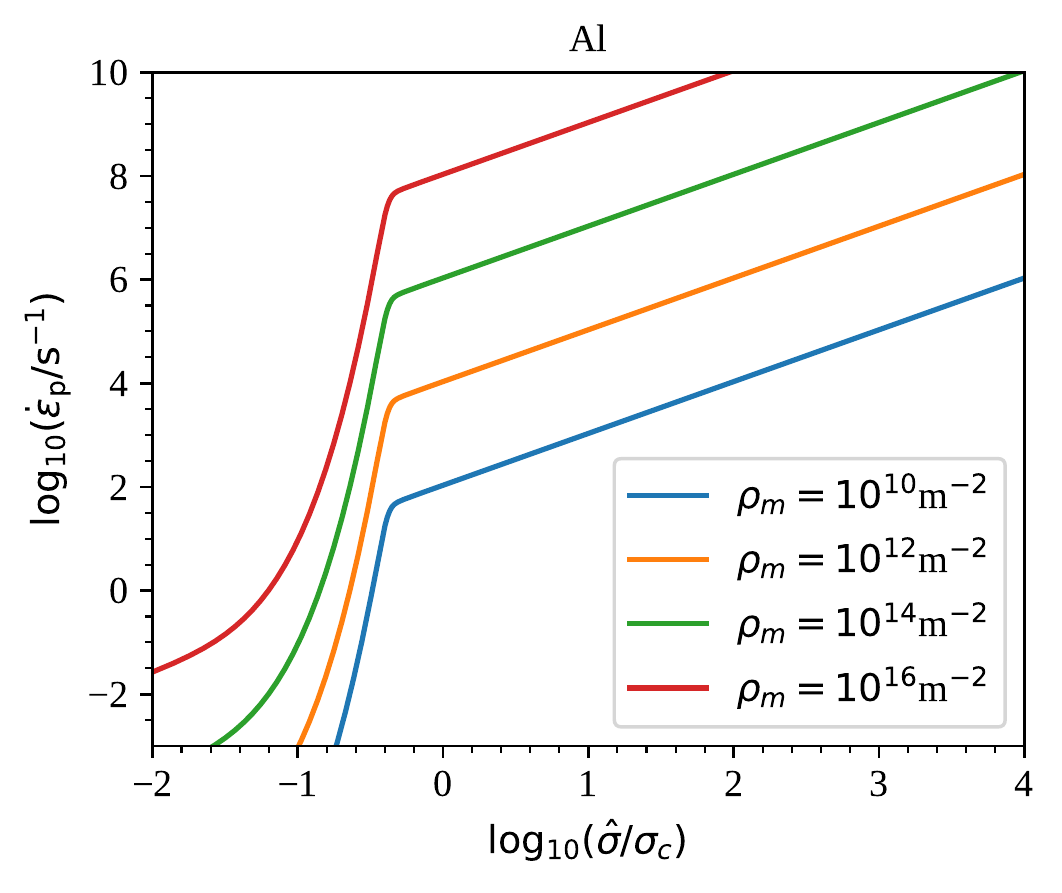}
 \includegraphics[width=0.5\textwidth]{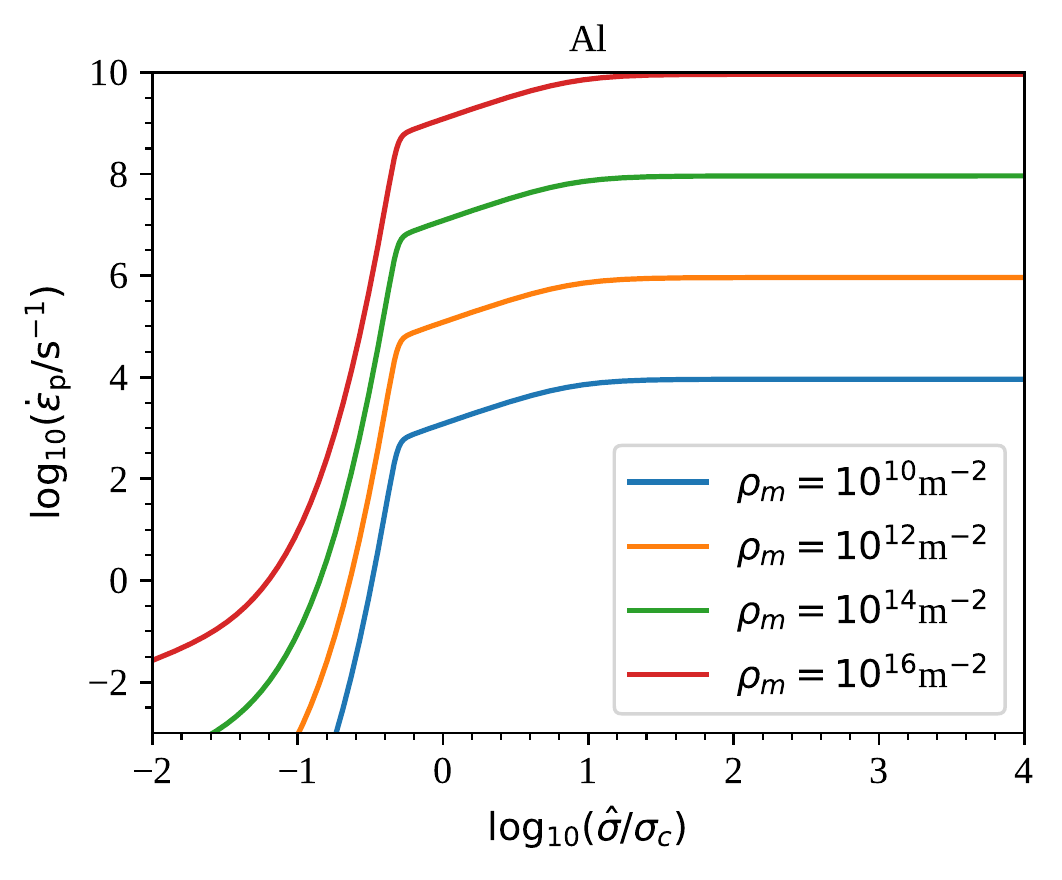}%
 \includegraphics[width=0.5\textwidth]{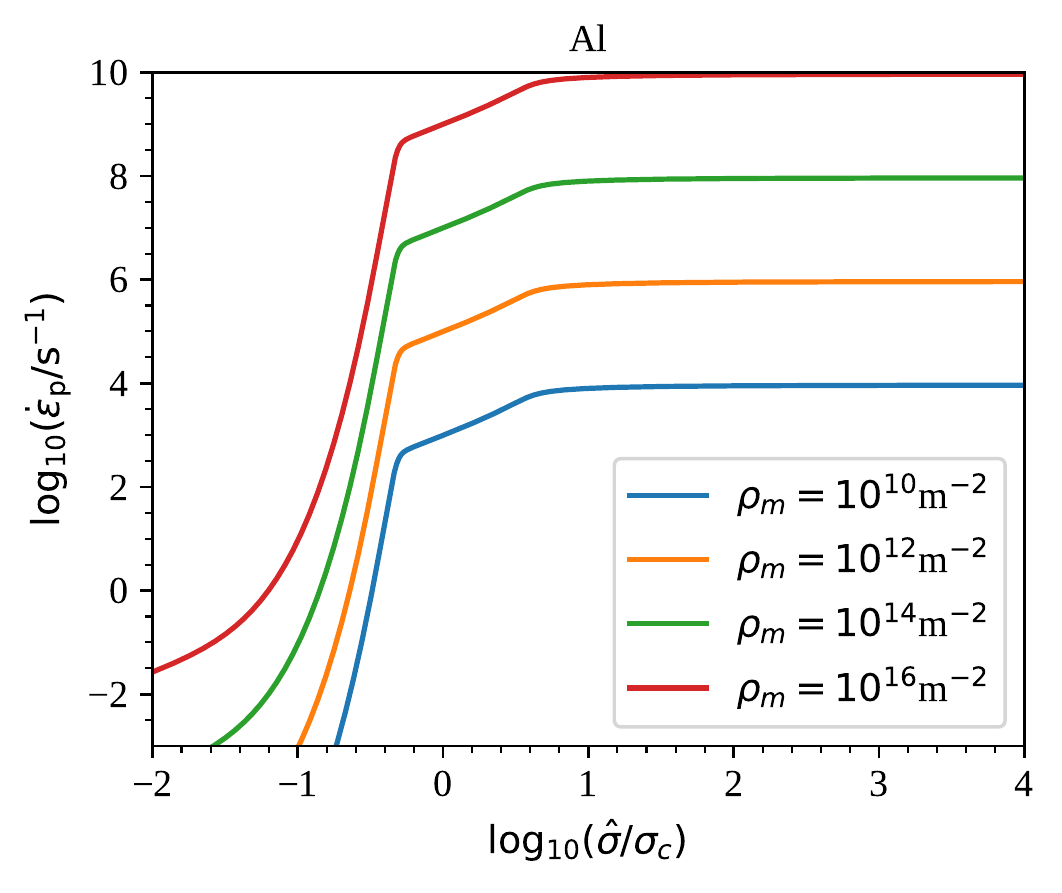}
 \caption{Mobile dislocation density dependence of $\dot\vare$ as given by the kinetic equation for aluminum for the case of a) constant drag coefficient $B=0.1$mPa\,s as used within the model of Ref.~\cite{Hunter:2015} (top left),
 b) drag coefficient $B(\hat\sigma=0,\rho,T)$ with $B_0=0.05$mPa\,s ($B_0$ at room temperature and ambient density) (top right),
 c) drag coefficient $B(\hat\sigma,\rho,T)$ in its simple functional form as given by \eqref{eq:simpleBofsigmaTrho} (bottom left),
 d) drag coefficient $B(\hat\sigma,\rho,T)$ in its full (numerically determined) form derived using \eqref{eq:fittedcurves} and \eqref{eq:vofsigma} (bottom right).
 Other parameters used in these figures are: $T=300$K, $\rho=\rho_0$ and $\rho_i=10^{12}$m$^{-2}$.}
 \label{fig:kinetic_Al_rhom}
\end{figure}
%%%%%%%%%%%%%%%%%%%%%%%%%%%%%%%%%%%%%%%%%%%%%%%%%%%%

\begin{figure}[!h!t]
 \centering
 \includegraphics[width=0.5\textwidth]{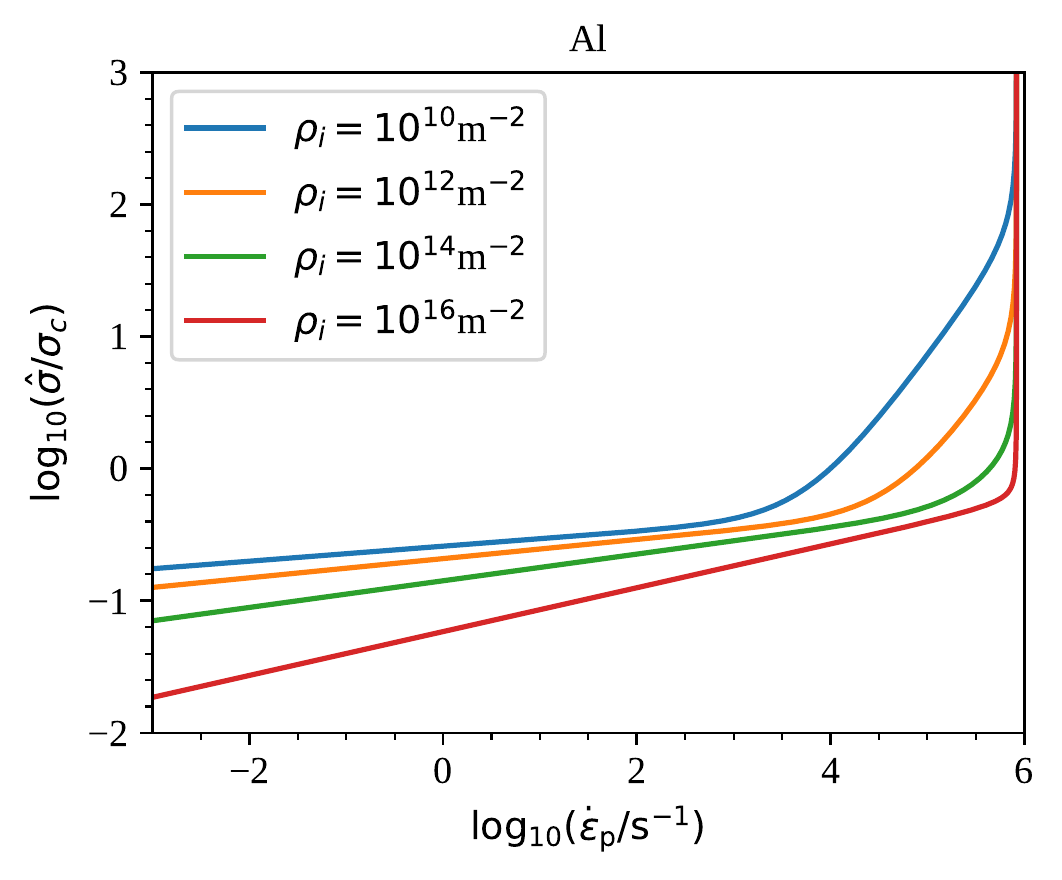}%
 \includegraphics[width=0.5\textwidth]{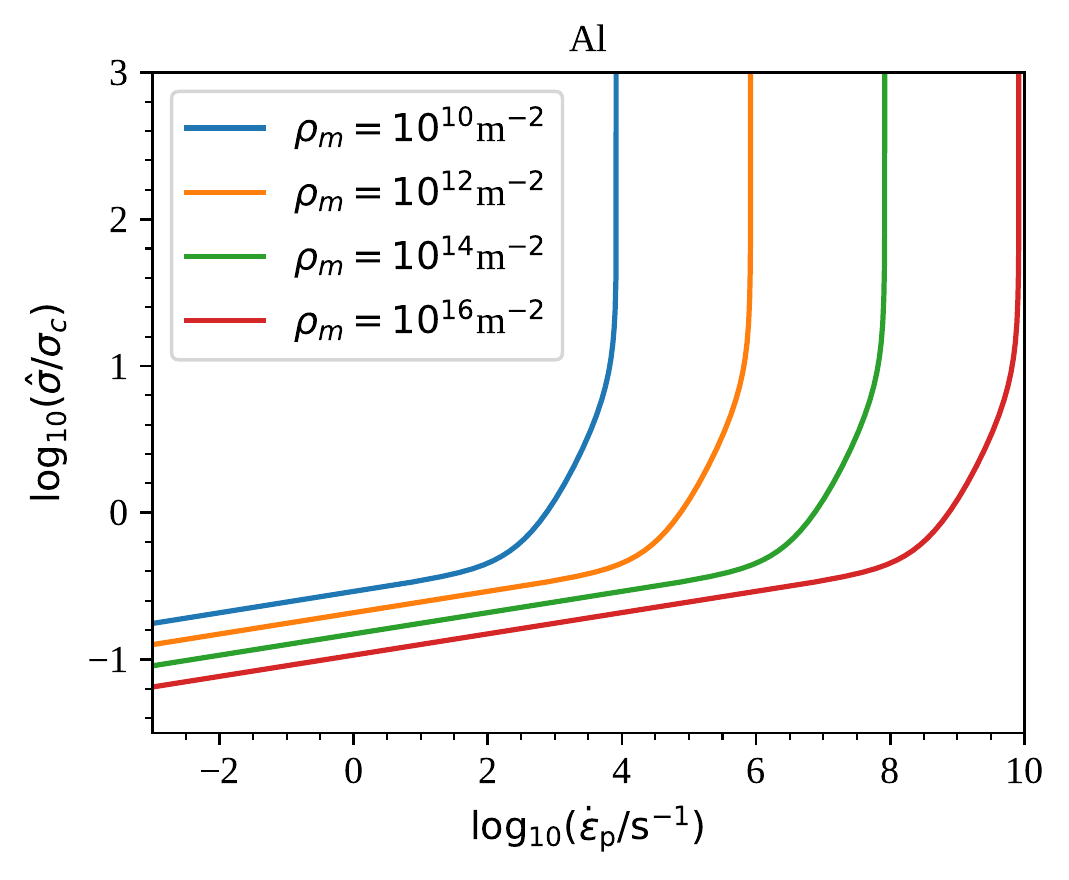}
 \includegraphics[width=0.5\textwidth]{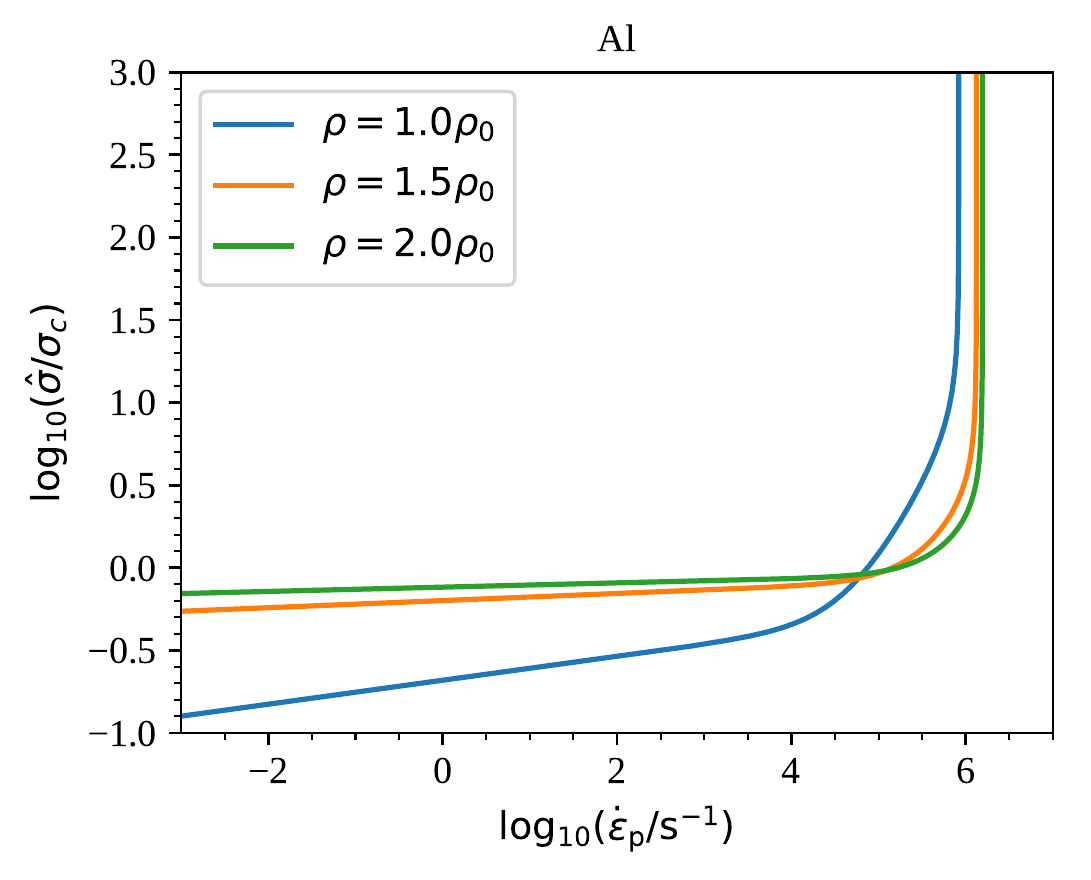}%
 \includegraphics[width=0.5\textwidth]{simple/inversekinetic_Al_T}
 \caption{Mobile/immobile dislocation density, material density, and temperature dependence of $\hat\sigma/\sigma_c$ as given by the inverse kinetic equation for aluminum for the case of drag coefficient $B(\hat\sigma,\rho,T)$ in its simple functional form as given by \eqref{eq:simpleBofsigmaTrho}.
 Large deviations to the kinetic equations shown in Figures \ref{fig:kinetic_Al_rho0}--\ref{fig:kinetic_Al_rhom} are clearly visible in the low stress regime and are due to the approximations being less accurate there.
 Unless otherwise stated in the figure legends, $T=300$, $\rho=\rho_0$, and $\rho_i=10^{12}$m$^{-2}=\rho_m$.}
 \label{fig:inversekinetic_simpleB_Al}
\end{figure}

\begin{figure}[!h!t]
 \centering
 \includegraphics[width=0.5\textwidth]{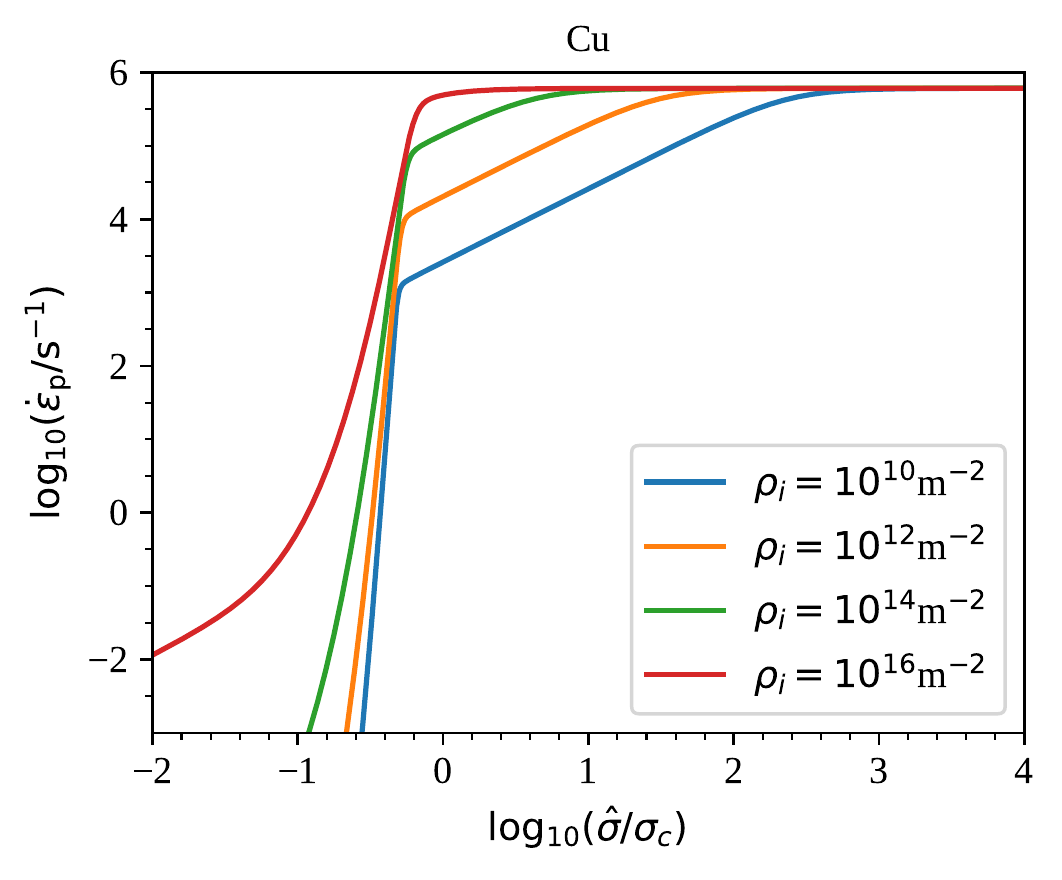}%
 \includegraphics[width=0.5\textwidth]{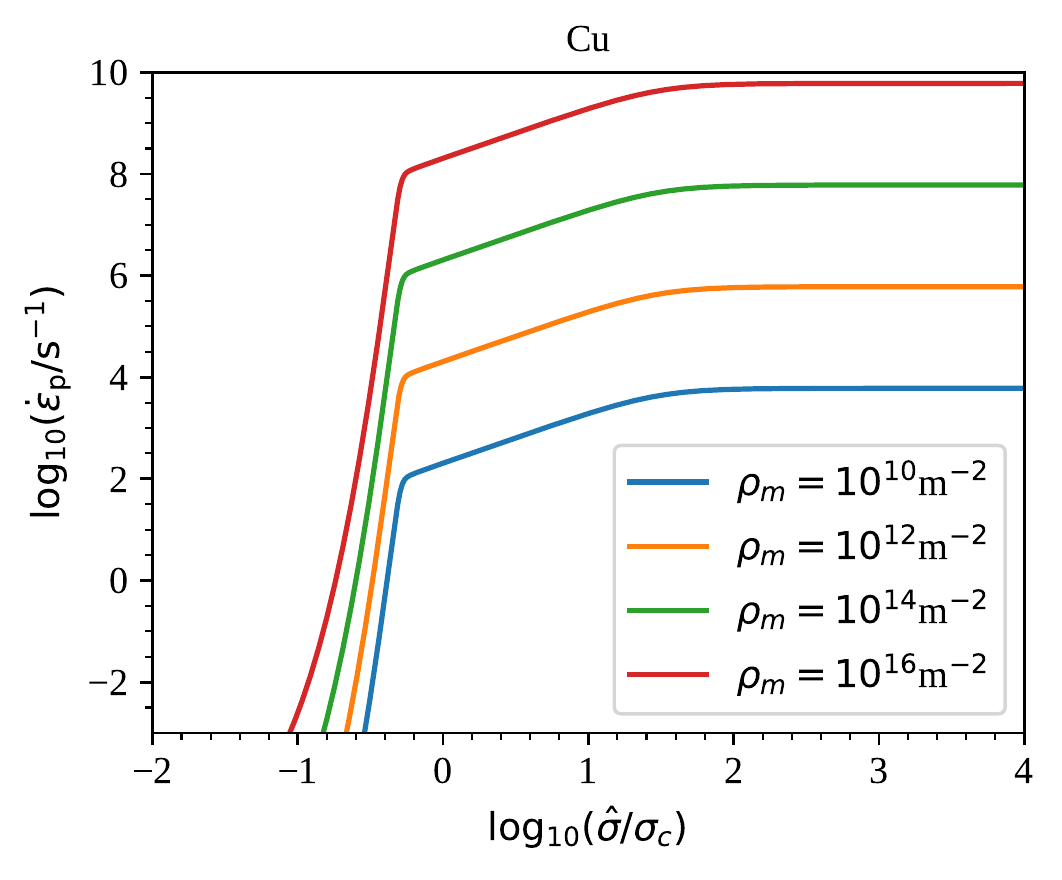}
 \includegraphics[width=0.5\textwidth]{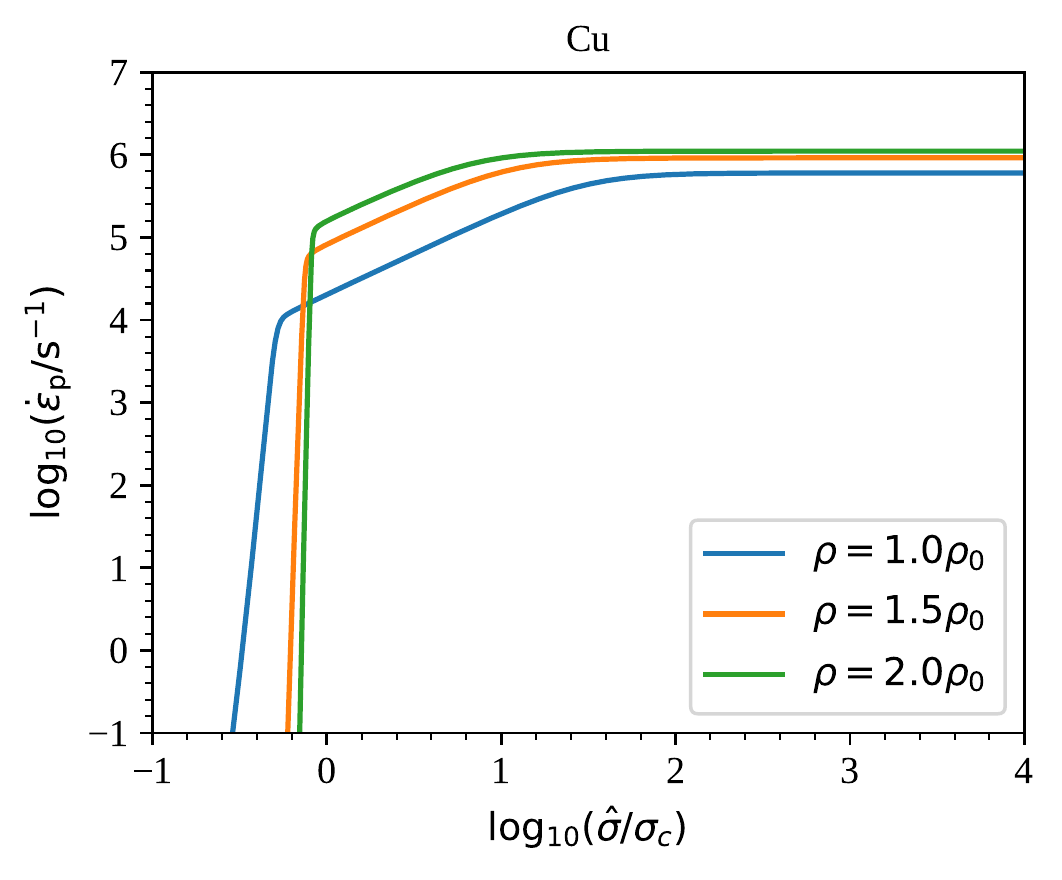}%
 \includegraphics[width=0.5\textwidth]{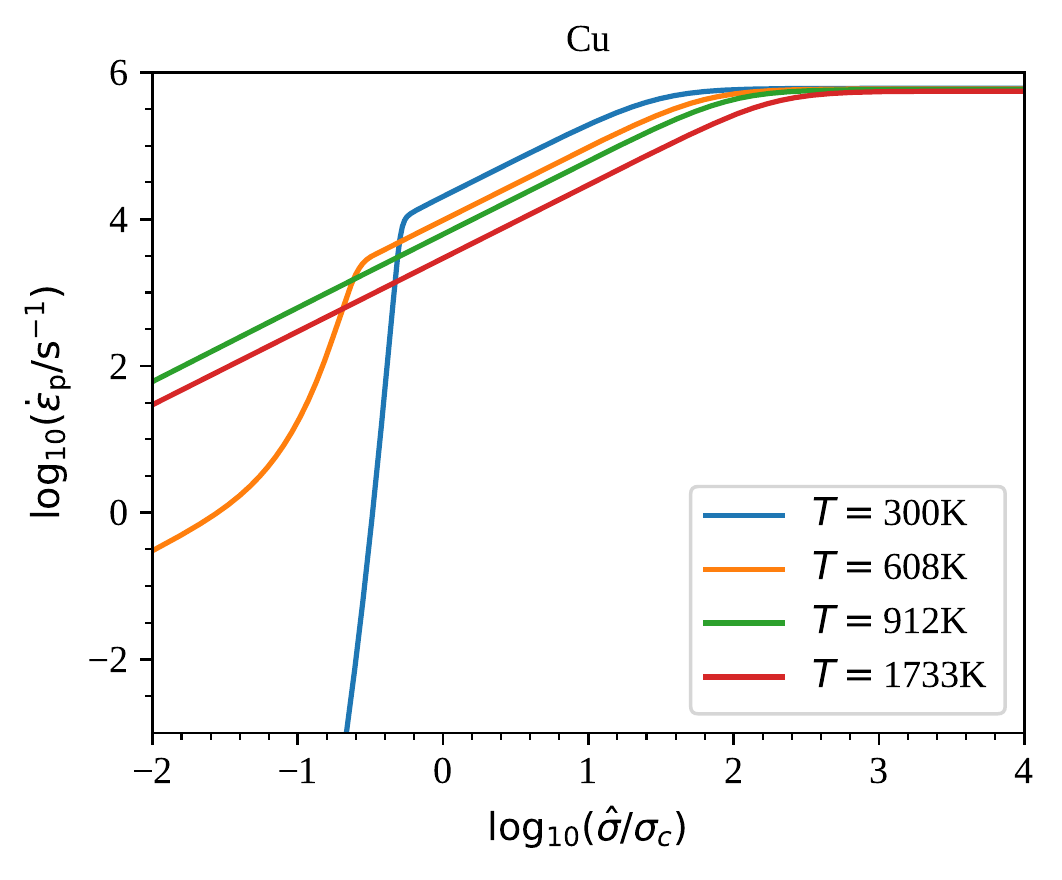}
 \caption{Mobile/immobile dislocation density, material density, and temperature dependence of $\dot\vare$ as given by the kinetic equation for copper for the case of drag coefficient $B(\hat\sigma,\rho,T)$ in its simple functional form as given by \eqref{eq:simpleBofsigmaTrho}.
 Unless otherwise stated in the figure legends, $T=300$, $\rho=\rho_0$, and $\rho_i=10^{12}$m$^{-2}=\rho_m$.}
 \label{fig:kinetic_simpleB_Cu}
\end{figure}

\begin{figure}[!h!t]
 \centering
 \includegraphics[width=0.5\textwidth]{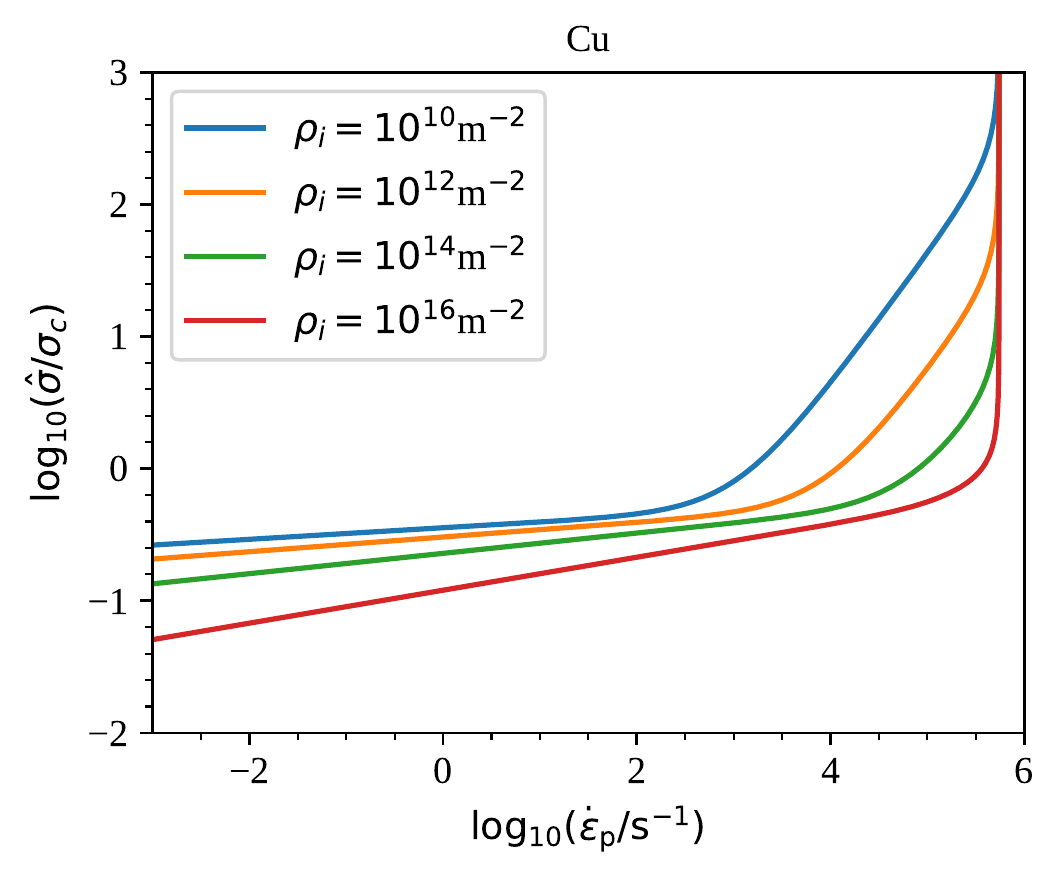}%
 \includegraphics[width=0.5\textwidth]{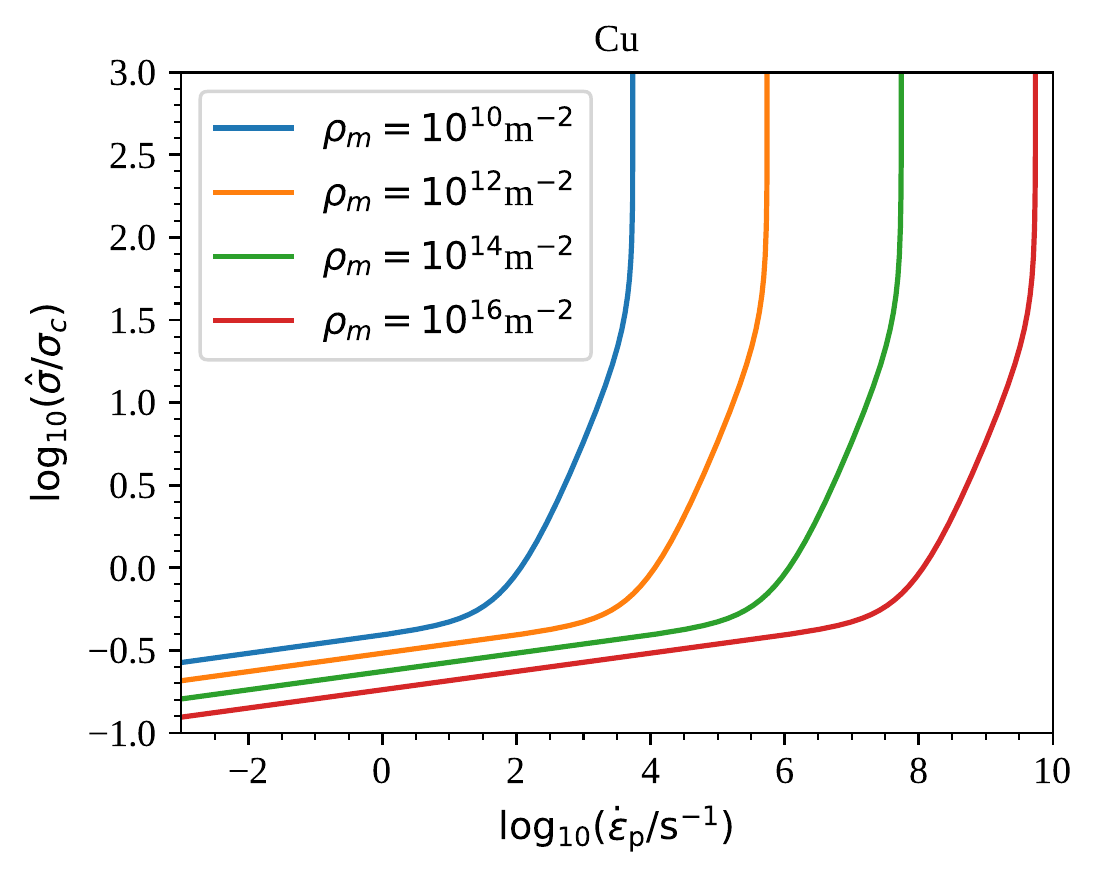}
 \includegraphics[width=0.5\textwidth]{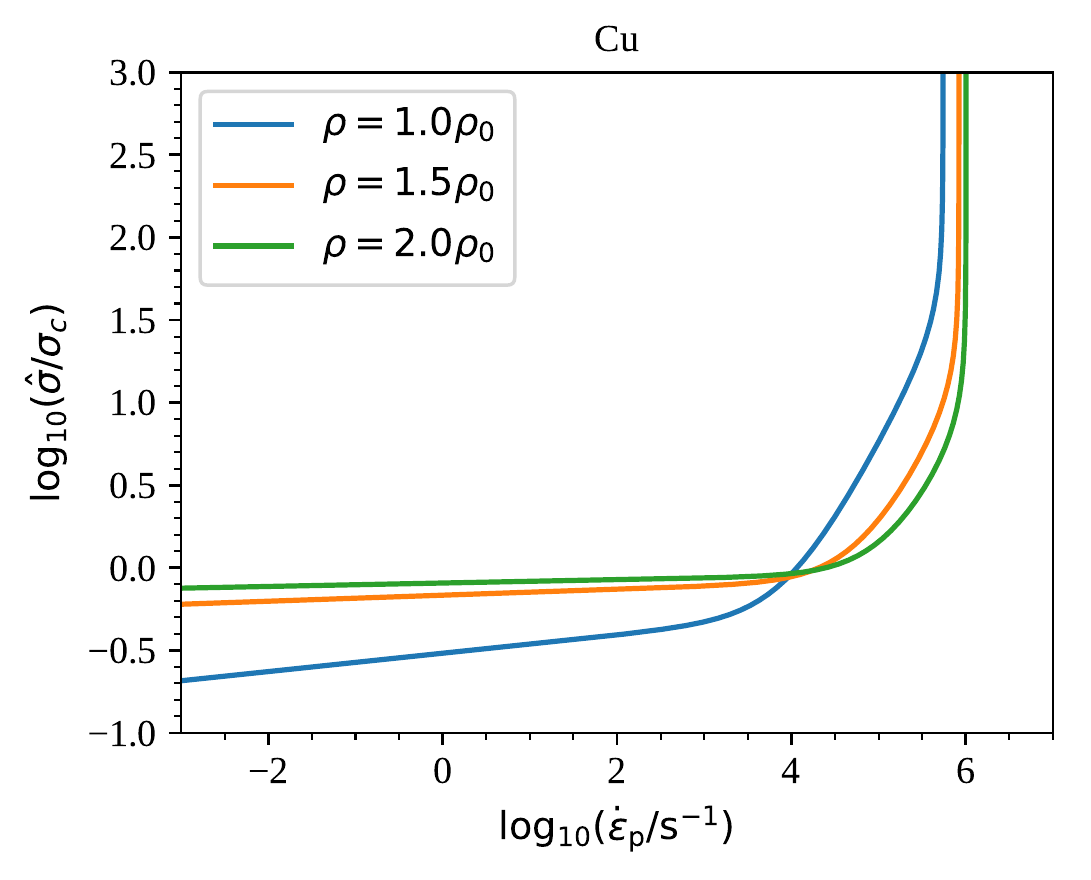}%
 \includegraphics[width=0.5\textwidth]{simple/inversekinetic_Cu_T}
 \caption{Mobile/immobile dislocation density, material density, and temperature dependence of $\hat\sigma/\sigma_c$ as given by the inverse kinetic equation for copper for the case of drag coefficient $B(\hat\sigma,\rho,T)$ in its simple functional form as given by \eqref{eq:simpleBofsigmaTrho}.
 Large deviations to the kinetic equations shown in Figure \ref{fig:kinetic_simpleB_Cu} are clearly visible in the low stress regime and are due to the approximations being less accurate there.
 Unless otherwise stated in the figure legends, $T=300$, $\rho=\rho_0$, and $\rho_i=10^{12}$m$^{-2}=\rho_m$.}
 \label{fig:inversekinetic_simpleB_Cu}
\end{figure}

Generalization of the kinetic equation \eqref{eq:kineticeqn} above to include $B\to B(\hat\sigma,\rho,T)$ is straightforward.
However, in order to generalize the inverse kinetic equation we need very simple analytic forms of $B$.
One straightforward generalization in this respect is $B\to B_0(\rho,T)$.
In order to take into account the more general \eqnref{eq:simpleBofsigmaTrho} for $B$, however, $b_+$ must be re-derived.
One finds
\begin{align}
 b_+ &= \log_{10}\left[\frac{\txt{s}^{-1}}{\rho_m b}v_0^{-1}\right]
 - \frac12\log_{10}\left[1 - \left(\frac{\txt{s}^{-1}v_0^{-1}\sigma_c 10^x}{(B_0 \ct \rho_m)}\right)^2\right]
 \,,\nn\\
 v_0^{-1} &= \left(\frac{2\phi_c}{\pi^2}+1\right)\frac{B_0}{b\sigma_c}
% \nn\\
% v_0^{-1} &= \left(\sqrt{\rho_i}\tilde{t}_B+\frac{B_0}{b\sigma_c}\right)
% \,,\qquad
% \tilde{t}_B = \frac{B_0\cL_i^2}{\pi^2\cT}
 \,, \label{eq:bplusnew}
\end{align}
i.e. the first term of $b_+$ is almost identical to \eqref{eq:defbplusold}, the only change being $B\to B_0(\rho,T)$ within
% both $t_B$ and
$v_0^{-1}$.
The second term introduces a strain rate (or $x$) dependence into $b_+$, which can be shown to be orders of magnitude smaller than the first term of \eqref{eq:bplusnew} in the interpolating regime close to $x_c$ allowing us to simply replace $b_+$ with this new expression within \eqref{eq:inversekinetic_paper}.
The $x$ dependent second term of $b_+$ only becomes important at high strain rates and ultimately diverges at a maximum strain rate of
\begin{align}
 \dot\vare_\txt{max}(\rho,T)&=\rho_m b(\rho)v_\txt{max}(\rho,T)
 \,, &
 v_\txt{max}(\rho,T)&=\frac{\ct(\rho,T)}{\left(\frac{2\phi_c}{\pi^2}+1\right)}
 \,, \label{eq:maxepsilon}
\end{align}
which has the form of Orowan's relation with effective maximum dislocation velocity $v_\txt{max}
%=\ct/\left(\frac{2\phi_c}{\pi^2}+1\right)
\approx 0.91\ct$.
The same maximum strain rate can of course be derived from \eqref{eq:kineticeqn} considering $\zeta\approx1$, $\sqrt{\rho_i}t_r \approx \sqrt{\rho_i}t_B\frac{\sigma_c}{\hat\sigma} + \sqrt{\rho_i}\tau ={2\phi_c B}/{\left(\pi^2 b\hat\sigma\right)} + \sqrt{\rho_i}\tau $ using \eqref{eq:estimates}, and \eqref{eq:simpleBofsigmaTrho} for $B$, which in this limit tends to $b\hat\sigma/\ct$.
Since $\sqrt{\rho_i}\tau \ll \frac1{\ct}\left(\frac{2\phi_c}{\pi^2}+1\right)$ even for the highest forest dislocation densities considered here (i.e. as long as $\rho_i\lesssim10^{16}$m$^{-2}$), this $\tau$-dependent term is negligible and we recover \eqnref{eq:maxepsilon}.

While the maximum dislocation velocity between obstacles (being limited by phonon drag) is the transverse sound speed, $\ct$, the effective highest velocity, $v_\txt{max}$, is reduced by the bow out time $t_B$ needed to overcome each obstacle, and $t_B$ in turn depends on $B$ in the current model.

\paragraph{Low strain rate regime revisited:}
%%%%%%%%%%%%%%%%%%%%%%%%%%%%%%%%%%%%%%%%%%%%%%

The inverse kinetic equation in the low stress/ low strain rate regime, \eqref{eq:bminus_old}, was derived assuming that the remobilization time was much larger than the run time.
However, this need not be the case when the temperature is large
since the associated increase in atomic vibrations enhances thermal activation.
Consequently the remobilization time may not be as long as initially thought, placing it closer in scale to the run time.
In addition, the increase in atomic vibrations can impact the drag (or run time) itself.
What this means for the inverse kinetic equation is that the interpolation point $x_c$, and hence the drag dominated regime, may get pushed to lower strain rates at sufficiently high temperature.

In order to quantify this effect, we revisit the kinetic equation \eqref{eq:kineticeqn}, and write its denominator on the r.h.s. as $\sqrt{\rho_i}t_r^\txt{eff}(\hat\sigma)+\frac{B(\hat\sigma)}{b\hat\sigma}=\sqrt{\rho_i}t_r^\txt{eff}(\hat\sigma)X(\hat\sigma)$.
Taking the limit $\hat\sigma\to0$ and using \eqref{eq:trlimit}, the correction factor, $X$, becomes
\begin{align}
 X(0)=1+\frac{4A B_0}{b\sigma_c\left[1+\erf(A)\right]\tau \sqrt{\pi\rho_i} e^{A^2}}
 \approx 1 + \frac{A B_0}{b\sigma_c \tau \sqrt{\rho_i}\, e^{A^2}}
 \,, \label{eq:X0}
\end{align}
with $B_0(\rho,T)$ given in \eqref{eq:simpleBofvTrho}.
This expression constitutes a lower bound for $X(\hat\sigma)$, and for wide parameter ranges, $X(0)\approx1$, but as temperatures rise, $A(T)$ drops and once $A\lesssim3.5$, the initially small correction can grow orders of magnitude greater than 1, i.e. the drag dominated regime persists down to lower stresses.
Unfortunately, this means that the approximations underlying the derivation of $m_-x+b_-$ break down, turning this regime into a non-trivial problem of the same complexity as if we were to invert the full kinetic equation.
We therefore take an alternative route to take $X>1$ into account in the inverse kinetic equation:
In particular, this effect can be accounted for by shifting the interpolation point $x_c$ by $-\log_{10}(X(0))$, thereby extending the regime where $m_+x+b_+$ is dominant.
We make the very conservative choice of $X(0)$ here in order to change the inverse kinetic equation as little as possible while still somewhat improving the approximation in the high temperature and low stress regime.
The only way to accommodate this shift in $x_c$ without affecting the high stress regime is by shifting $b_-$ accordingly, see \eqref{eq:inversekinetic_paper}.
Note, that we are not including drag effects in $m_-x+b_-$, rather we ensure $m_+x+b_+$ is dominant whenever drag effects are important by further sacrificing accuracy (at high $T$) in a regime where the approximations leading to $m_-x+b_-$ break down anyway (i.e. when $y<-1$).
Hence our new $b_-$ presently reads
\begin{align}
 b_- &= -m_-\log_{10}\left(\frac{\rho_m b}{\sqrt{\rho_i}\tau\txt{s}^{-1}}\right) - (m_+ - m_-)\log_{10}\left(1 + \frac{A B_0}{b\,\sigma_c \tau\sqrt{\rho_i}\, e^{A^2}}\right)
 \,, \label{eq:bminusnew}
\end{align}
and Figure \ref{fig:bminus} shows a comparison of old and new $b_-$ for Cu and Al assuming $\rho=\rho_0$ and $\rho_i=10^{12}$m$^{-2}=\rho_m$.
Thus, in the high temperature regime, the new constant $b_-$ shifts the interpolation point $x_c$ down to negative values expanding the drag dominated regime, see \eqref{eq:inversekinetic_paper}.
The effect of the new low strain rate formulation and a non-constant drag coefficient can be seen in comparison to the originally developed kinetic equation in Figure \ref{fig:inversekinetic_Tdep}.
Several differences are obviously apparent, perhaps most notable are the differences in the intermediate strain rate regime, where the curves now `split' due to the temperature dependence in $B$, and the existence of a maximum achievable strain rate due to the divergence of $B$ at a maximum dislocation velocity.
Note that this maximum strain rate can be significantly higher if the density of mobile dislocations is increased; see \eqnref{eq:maxepsilon} and Figure \ref{fig:kinetic_Al_rhom}.
In addition, the low strain rate regimes of Figure \ref{fig:inversekinetic_Tdep} show how the refinement of $ b_-$ improves the inverse kinetic equation, i.e. the low strain regime is closer to what it should be in comparison to the kinetic equation.
In the following section and Figures \ref{fig:kinetic_Al_rho0}--\ref{fig:inversekinetic_simpleB_Cu} we discuss the results of our present improvements in more detail.

\section{Results from the integrated model}
\label{sec:results}
%%%%%%%%%%%%%%%%%%%%%%%%%%%%%%%%%%

Figures \ref{fig:kinetic_Al_rho0}--\ref{fig:inversekinetic_simpleB_Cu} compare the various cases of kinetic and inverse kinetic equations discussed above for Al and Cu.
The present model is an isotropic one, and hence can be expected to be applicable especially for (fcc) metals which are fairly isotropic in the sense that their Zener ratio is close to 1.
We therefore first elaborate on the (inverse) kinetic equation for aluminum, which has a Zener ratio of $1.22$, making it more isotropic then copper (whose Zener ratio is $3.21$).
Nonetheless, we present results also for the latter in this section and in Figures \ref{fig:kinetic_simpleB_Cu}, \ref{fig:inversekinetic_simpleB_Cu}.
In particular, we compared the effects of four different drag coefficients in Figures \ref{fig:kinetic_Al_rho0}--\ref{fig:kinetic_Al_rhom}:
\begin{enumerate}
\itemsep=0pt
 \item $B=0.1$mPas (and constant),
 \item $B=B_0(\rho,T)$ (and independent of stress),
 \item the simple functional form $B(\hat\sigma,\rho,T)$ of \eqnref{eq:simpleBofsigmaTrho},
 \item the full numerically determined drag coefficient $B(\hat\sigma,\rho,T)$ computed from \eqref{eq:vofsigma} with \eqref{eq:fittedcurves}, $B=(B_\txt{e}+B_\txt{s})/2$, and using density and temperature scaling of $B(v)$, $b$, and $\ct=\sqrt{G/\rho}$ as given in \eqref{eq:Bscaling},
\eqref{eq:densityscaling}, and \eqref{eq:shearmod}.
\end{enumerate}
The only visible difference between the latter two cases is seen around $1\lesssim\log_{10}(\hat\sigma/\sigma_c)\lesssim1.5$, where the curves are slightly smoothed out by the simple functional form for $B$,
see the bottom rows of Figures \ref{fig:kinetic_Al_rho0}--\ref{fig:kinetic_Al_rhom}, but especially Fig. \ref{fig:kinetic_Al_T} where we have highlighted this tiny effect using inset plots.
This can be traced back to averaging out the initial drop in $B$ seen in Figure \ref{fig:compareBsimple} which is well within the uncertainty of $B$.
Hence, the present comparison confirms that \eqnref{eq:simpleBofsigmaTrho} is a good approximation to $B(\sigma)$ within the kinetic equation.
For the inverse kinetic equation (Figures \ref{fig:inversekinetic_simpleB_Al} and \ref{fig:inversekinetic_simpleB_Cu}) we made use of \eqref{eq:bminusnew}.
Unless otherwise stated in the figure legends, we considered $T=300$, $\rho=\rho_0$, and $\rho_i=10^{12}$m$^{-2}=\rho_m$ in all of these figures.

The maximum strain rate \eqref{eq:maxepsilon} is clearly visible in the bottom rows of Figures \ref{fig:kinetic_Al_rho0}--\ref{fig:kinetic_Al_rhom} as well as in the inverse kinetic equation shown in Figures \ref{fig:inversekinetic_simpleB_Al} and \ref{fig:inversekinetic_simpleB_Cu}, constituting the most notable difference to the earlier results of \cite{Hunter:2015} shown on the top left of those figures.
The only case where a $\tau$-dominated maximum strain rate of $\dot\vare_\txt{max}(\rho,T)=\rho_m b(\rho)/(\sqrt{\rho_i}\tau)$ is seen, is for $\rho_i=10^{16}$m$^{-2}$, i.e.,
% and for $\rho=2\rho_0$, i.e.,
the time scale $\tau$ is only important in the highest stress regime when the stress dependence of $B$ is neglected and simultaneously
% either forest dislocation density or material density are extremely high.
the forest dislocation density is extremely high.
In this respect we must stress, that the (im)mobile dislocation densities need to be determined dynamically, and a density evolution model is currently being developed \cite{Hunter:wip}.
For the purpose of our present study, we vary both densities within the ranges $10^{10}\mathrm{m}^{-2}\le\rho_{i,m}\le10^{16}\mathrm{m}^{-2}$, the highest value being comparable to the highest experimentally measured dislocation densities (produced by overdriven shocks); see \cite{Meyers:2003,Meyers:2009} and references therein.

Other visible differences to the earlier results of \cite{Hunter:2015} that are seen in Figures~\ref{fig:kinetic_Al_rho0}--\ref{fig:kinetic_Al_rhom} are summarized as follows:
The density dependence of $B$ leads to an enhanced splitting of the curves in the drag dominated regime, as shown in Figure \ref{fig:kinetic_Al_rho0}:
The top left sub-figure is the only one neglecting the density dependence of $B$.
The top right sub-figure shows the enhanced splitting most clearly, but even in the bottom row it is visible in the drag dominated regime before approaching the maximum strain rate.
Also, the temperature dependence of $B$ leads to a splitting of the curves, as seen in Figure \ref{fig:kinetic_Al_T}.
Figures \ref{fig:kinetic_Al_rhoi} and \ref{fig:kinetic_Al_rhom} show that the inclusion of $B(\hat\sigma,\rho,T)$ does not cause any noticeable difference in the separation of these curves.
Finally, the generalized $b_-$ of the inverse kinetic equation, \eqref{eq:bminusnew}, slightly extended the drag dominated regime down to lower strain rates for temperatures approaching melting, thus matching more accurately the kinetic equation at high temperatures and sufficiently large local stress, i.e. $\log_{10}(\hat\sigma/\sigma_c)\gtrsim-1$.
At lower stress, $\log_{10}(\hat\sigma/\sigma_c)\ll-1$ the linear approximation $y\approx m_-x+b_-$ of \eqref{eq:bminus_old} (with \eqref{eq:bminusnew}) breaks down and hence the inverse kinetic equation deviates visibly from its more accurate counter part in this regime.
Note, however, that even the kinetic equation presented here cannot be used at very low stress, since other effects, e.g. creep, which we do not discuss here, become important.

\section{Conclusions}
\label{sec:conclusions}
%%%%%%%%%%%%%%%%%%%%%%%%%%%%%%%%%%

We have incorporated a model for the dependence of the dislocation drag coefficient $B(\hat\sigma,\rho,T)$ (developed using first principle calculations) into a plastic constitutive model, which has generalized the van 't Hoff-Arrhenius low strain rate relationship up to strain rates of roughly  $10^{10} \mathrm{s}^{-1}$, allowing for the effects of dislocation drag on the overall material response in the high-rate loading regime to be studied.
In doing so, we have derived an approximate analytical form for the drag coefficient $B(\hat\sigma,\rho,T)$ as a function of stress (or velocity), material density, and temperature, which can be computed in-line within the plastic constitutive model.
Since plastic constitutive models are generally expected to determine the flow stress as a function of strain rate (rather than strain rate as a function of stress), we additionally presented approximations for the inverse of our generalized kinetic equation.

We subsequently demonstrated the importance of taking into account a realistic drag coefficient in its full functional form, as important features of flow stress versus strain rate in the drag dominated regime would be missed otherwise.
In particular, these features are:
\begin{itemize}
\itemsep=0pt
 \item the temperature dependence shown in Figure \ref{fig:kinetic_Al_T},
 \item the enhanced material density dependence shown in Figure \ref{fig:kinetic_Al_rho0},
 \item and the maximum strain rate at given mobile dislocation density (Figure \ref{fig:kinetic_Al_rhom}).
\end{itemize}
As for the last point, we must stress that in order to derive a true maximum strain rate, a model of mobile/immobile dislocation density evolution to determine an upper bound on the mobile dislocation density is also required.
The latter is work in progress~\cite{Hunter:wip}.
An upper limit on the mobile dislocation density in conjunction with a limiting dislocation velocity yields an upper limit on the strain rate of a given material and in comparing to the highest experimentally determined strain rate, one could shed some light on the open question whether supersonic dislocations exist or not.
At this point, however, all we can state is that the highest experimental strain rates of $10^{10} \mathrm{s}^{-1}$ are consistent with the maximum model strain rate when dislocation velocities are limited by the transverse sound speed and when the density of mobile dislocations is comparable to the highest experimentally determined total dislocation density  (produced by overdriven shocks \cite{Meyers:2003,Meyers:2009}) of about $10^{16} \mathrm{m}^{-2}$; see Figures \ref{fig:kinetic_Al_rhom} and \ref{fig:kinetic_simpleB_Cu}.

An important conclusion of this work that was previously missed is that the drag dominated regime can extend down to lower strain rates ($\sim10^{2-4}$s$^{-1}$, see Figures \ref{fig:kinetic_Al_T}, \ref{fig:kinetic_simpleB_Cu}) if the temperature is sufficiently high (i.e. above roughly half the melting temperature).
The reason for this feature is that the drag coefficient grows roughly linearly with temperature whereas thermal activation is significantly enhanced because the remobilization time $t_r$ at obstacles decreases exponentially with temperature (via a van 't Hoff-Arrhenius law) due to the associated increase in atomic vibrational amplitudes.
Therefore, $t_r$ becomes comparable or even smaller than the dislocation run time due to phonon drag, even at moderate stresses if the temperature is high (see Figures \ref{fig:kinetic_Al_T} and \ref{fig:kinetic_simpleB_Cu}).
This observation emphasizes the importance of taking into account dislocation drag in its full functional form even at moderate strain rates if the temperature is sufficiently high.

\subsection*{Acknowledgements}
%%%%%%%%%%%%%%%%%%%%%%%%%%%%%%%%%%%%%%%%%%
\noindent
This work was performed under the auspices of the U.S. Department of Energy under contract 89233218CNA000001.
In particular, the authors are grateful for the support of the Materials project within the Advanced Simulation and Computing, Physics and Engineering Models Program.
D.~N. B. thanks S.~J. Fensin, R.~G. Hoagland, and D.~J. Luscher for related interesting discussions.

% \clearpage
%%%%%%%%%%%%%%%%%%%%%%%%%%%%%%%%%%%%%%%%%%%
\bibliographystyle{utphys-custom}
\bibliography{dislocations}
%%%%%%%%%%%%%%%%%%%%%%%%%%%%%%%%%%%%%%%%%%%

\end{document}